\documentclass[ba,preprint]{imsart}
\pubyear{2023}
\volume{TBA}
\issue{TBA}
\firstpage{1}
\lastpage{1}

\usepackage{amsthm}
\usepackage{amsmath}
\usepackage{amssymb}
\usepackage{natbib}
\usepackage[colorlinks,citecolor=blue,urlcolor=blue,filecolor=blue,backref=page, bookmarks=false]{hyperref}
\usepackage{graphicx}
\usepackage{microtype} 

\usepackage{bm}
\usepackage{amssymb,latexsym, amsfonts, amscd, xy}
\usepackage{booktabs}
\usepackage[shortlabels]{enumitem}
\usepackage[export]{adjustbox}

\usepackage[boxed]{algorithm2e}

\startlocaldefs

\newcommand{\reals}{\mathbb{R}}
\newcommand\given{\,|\,}
\newcommand\zb{\bm z}
\newcommand\xb{\bm x}
\newcommand\xib{\bm \xi}
\newcommand\pib{\bm \pi}
\newcommand\etab{\bm \eta}
\newcommand{\nats}{\mathbb{N}}
\newcommand\norm[1]{\|#1\|}
\newcommand\ct{\widetilde c}
\newcommand\mask{m}
\newcommand\maskb{\bm m}

\def\R{\mathbb{R}}

\DeclareMathOperator\Dir{\mathsf{Dir}}
\DeclareMathOperator{\CRP}{\mathsf{CRP}}
\DeclareMathOperator{\Beta}{\mathsf{Beta}}

\newcommand{\casc}{\textsc{casc}}
\newcommand{\bcdc}{\textsc{bcdc}}
\newcommand{\bsbm}{\textsc{bsbm}}
\newcommand{\true}{\textsc{true}}
\newcommand{\cascore}{\textsc{cascore}}
\newcommand{\spec}{\textsc{sc}}
\newcommand{\kmeans}{$k$-{\textsc{means}}}

\usepackage{xcolor}

\newtheorem*{remark}{Remark}
\newtheorem*{example}{Example}

\endlocaldefs

\begin{document}


\begin{frontmatter}
\title{Bayesian community detection for networks with covariates}
\runtitle{BCDC}

\begin{aug}
\author{\fnms{Luyi} \snm{Shen}\thanksref{addr1,m1}\ead[label=e1]{lshen4@nd.edu}},
\author{\fnms{Arash} \snm{Amini}\thanksref{addr2,m2}\ead[label=e2]{aaamini@ucla.edu}}
\author{\fnms{Nathaniel} \snm{Josephs}\thanksref{addr3,m3}\ead[label=e3]{nathaniel.josephs@yale.edu}}
\and
\author{\fnms{Lizhen} \snm{Lin}\thanksref{addr1,m1}%
\ead[label=e4]{lizhen.lin@nd.edu}}

\runauthor{L. Shen et al.}

\address[addr1]{Department of Applied and Computational Mathematics and Statistics, The University of Notre Dame
    \printead{e1} 
    \printead*{e4}
}

\address[addr2]{Department Statistics, UCLA
    \printead{e2}
}

\address[addr3]{Department of Biostatistics, Yale School of Public Health
    \printead{e3}
}

\end{aug}

\begin{abstract}
	The increasing prevalence of network data in a vast variety of fields and the need to extract useful information out of them have spurred fast developments in related models and algorithms.
	Among the various learning tasks with network data, community detection, the discovery of node clusters or ``communities," has arguably received the most attention in the scientific community.
	In many real-world applications, the network data often come with additional information in the form of node or edge covariates that should ideally be leveraged for inference.
	In this paper, we add to a limited literature on community detection for networks with covariates by proposing a Bayesian stochastic block model with a covariate-dependent random partition prior.
	Under our prior, the covariates are explicitly expressed in specifying the prior distribution on the cluster membership.
	Our model has the flexibility of modeling uncertainties of all the parameter estimates including the community membership.
	Importantly, and unlike the majority of existing methods, our model has the ability to learn the number of the communities via posterior inference without having to assume it to be known.
	Our model can be applied to community detection in both dense and sparse networks, with both categorical and continuous covariates, and our MCMC algorithm is very efficient with good mixing properties.
	We demonstrate the superior performance of our model over existing models in a comprehensive simulation study and an application to two real datasets.
\end{abstract}


\end{frontmatter}

	
\section{Introduction}
	
The ubiquity of network data in modern science and engineering and the need to extract meaningful information out of them has spurred rapid developments in the models, theory, and algorithms for the inference of networks \citep{1959erdos, bickel2009nonparametric, graphon1, kolaczyk2009statistical, lovasz2012large, eric-aos}. Among the specific learning tasks with network data, community detection, which aims to detect communities or clusters among nodes, has arguably received the most attention in the scientific community. Various models and algorithms have been developed for community detection in networks including modularity-based methods \citep{Newman06}, spectral clustering algorithms \citep{LuxburgTutorial07,rohe_yu}, stochastic block models \citep{holland1983stochastic, newmanblock,ball2011efficient}, optimization-based approaches via semidefinite programming \citep{amini2014semidefinite}, and various Bayesian models \citep{BCD2012, hsbm}, among others.  
	
\begin{figure}
	 \centering
	 \includegraphics[width=.75\textwidth]{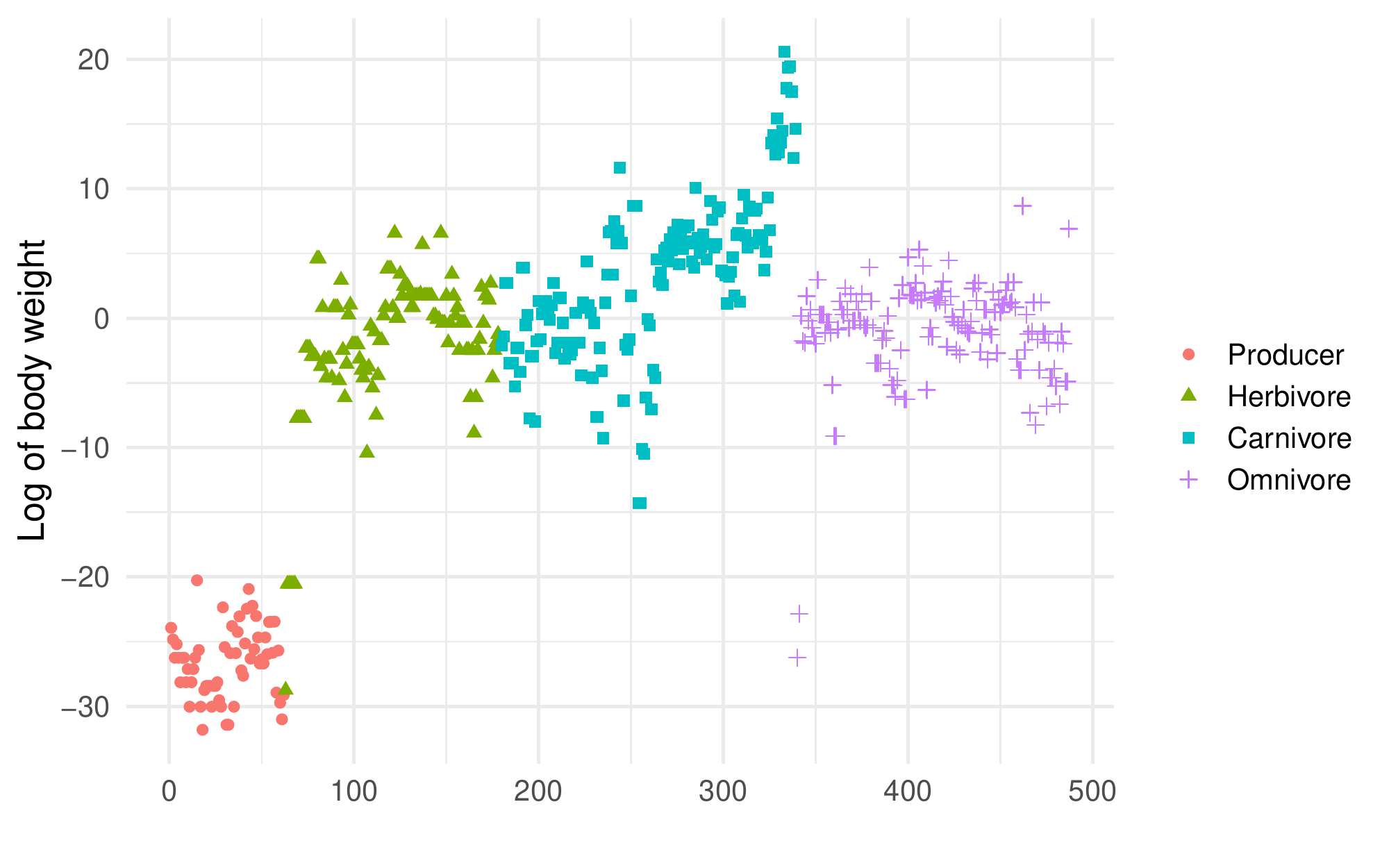}
	 \caption{Log body mass for different species colored by their feeding type.}
	 \label{fig:bodymass}
\end{figure}

Besides the edge information of an observed network, there are often additional covariates or nodal information available in many real-world networks.
These additional covariates should be ideally utilized when performing community detection.
For example, in a Facebook network, one can obtain from an individual's profile covariates including current city, workplace, hometown, education, and hobbies.
Another example is the Weddell Sea trophic network, which describes the marine ecosystem of the Weddell Sea \citep{jacob2011role}.
It is a predator-prey network that includes the average adult body mass for each of the species.
If one were to only utilize the network information, it is hard to differentiate all the different feeding types.
However, the body mass of each species shows a partial clustering by the group, as seen in Figure \ref{fig:bodymass}.
Therefore, a better clustering should be achievable when both the network and covariates are incorporated.

Such network data have motivated an emerging line of work that aims to deal with community detection problems that leverage both the network and the exogenous covariates.
A node-coupled SBM is proposed in \cite{weng2016community} in which cluster information or the block matrix is uniquely encoded by the covariates.
Another model from \cite{zhang2017node} specifies that the link probability between a pair of nodes is contributed additively by the block probability in an SBM and a similarity measure between the covariates of a pair of nodes.
A similar class of block models is proposed in \cite{sweet2015} that also accommodates covariates in an additive way such that the link probability is influenced by both block membership and covariates.
A covariate-assisted spectral clustering algorithm is proposed in \cite{binkiewicz2017covariate} and later modified for degree-corrected block models in \cite{hu2022covariate}.
Categorical covariates on the actor level are included in the model in \cite{tallberg2004}, and the block affiliation probabilities are modeled conditional on the covariates via a multinomial probit model.
Another prominent method in the frequentist literature is due to \cite{zhang2016community} in which a joint community detection criterion is proposed using both the adjacency matrix of the network and the node features, and their algorithm weights the edges according to feature similarities.
Recently, the interplay between network information and covariates is investigated in an optimization framework for community detection in sparse networks in \cite{sarkar}.
From a Bayesian perspective, there are a few papers that are closely related to our work. We introduce them in Section~\ref{sec:comparison} and provide a detailed discussion of their connections to our work (and to each other).

We add to this literature by proposing a Bayesian community detection procedure in which the effects of the covariates are incorporated via a covariate-dependent random partition prior on the node labels of an SBM.
The covariates are explicitly expressed and incorporated in the prior probability of generating clusters.
One of the distinctive features of our models compared with the ones already proposed in the literature is that ours has the ability to learn the number of communities via posterior inference without having to assume it to be known.
The proposed model has the flexibility of assessing uncertainties for all the model parameters through an efficient MCMC algorithm for posterior inference.
Note that there are several works in the literature that have employed the idea of a random partition prior or Bayesian nonparametric models for modeling network or relational data.
We discuss these comparisons in Section~\ref{sec:comparison} after introducing our model.

Our model can be applied in both dense and sparse regimes.
In a sparse regime, as one of our simulation studies shows, our model outperforms other state-of-the-art methods such as that of \cite{sarkar}, whose primary goal was to deal with sparse network condition with covariates.
We also apply our methods to networks that have covariates with relatively high-dimensions.
Our extensive simulations demonstrate our overall superior performance over existing methods in networks with continuous or categorical features, even when those methods are given the true number of communities.

The remainder of our paper is organized as follows.
Section \ref{sec:model} is devoted to our model description and MCMC algorithms.
In Section \ref{sec:simu}, we carry out several simulation studies in various settings to demonstrate the utility of our proposed model and algorithms.
We also apply our model to two data examples in Section \ref{sec:data}.
We conclude in Section \ref{sec:discussion} with possiblities for future work.
	
\section{Prior, model, and MCMC algorithm} \label{sec:model}
	
Consider an observed network on $n$ nodes represented by an $n\times n$ adjacency matrix $A = (A_{ij})$ with $A_{ij}=1$ indicating the presence of a link between nodes $i$ and $j$, and $A_{ij}=0$ otherwise.
Assume in addition that we have some covariate information $x_i \in \reals^p$ for each node $i = 1,\dots,n$.
The covariate information associated with the node are often referred to as nodal information or node features of the network, and are frequently encountered in modern network data.
We let $\xb=(x_1,\ldots, x_n)^T \in~\R^{n\times p}$ denote all of the node covariates of a given network.
Our goal is to perform network community detection by incorporating both the network structure and the nodal information. 
The key challenge is how to jointly model  these two sources of information. 
Below we propose a Bayesian model that incorporates the nodal information in the prior probability of cluster membership within an SBM.
	
Let $\zb=(z_1,\ldots, z_n) \in \nats^n$ be a node membership vector and $L(\zb) = \max\{z_i: i \in [n]\}$ indicate the total number of clusters implied by $\zb$.
We do not assume $L(\zb)$ to be known \textit{a priori}.
Let $S_\ell(\zb)=\{i\in[n]:\, z_i=\ell\}$ be the set of indices of nodes belonging to the $\ell^{th}$ cluster according to $\zb$.
For any subset $S \subseteq [n]$ and $\xb=(x_1,\cdots,x_n)^T \in~\reals^{n\times p}$, let $g(S\given \xb)$ be a nonnegative function that measures the homogeneity of the covariates $\{x_i,~i \in S\}$.
That is, $g(S \given \xb)$ takes larger values when all of the $x_i$ with $i \in S$ are more similar.
One can think of $S$ as a potential cluster of nodes and $g(S \given \xb)$ as a measure of the quality of such cluster, with regards the nodal information 

Inspired by~\cite{muller2010random, park2010bayesian, muller2011product}, we consider the following covariate-dependent random partition model:
\begin{equation} \label{2_1}
    p(\zb \given \xb)\propto \prod^{L(\zb)}_{\ell=1} g\big(S_\ell(\zb) \given \xb\big) \cdot c\big(S_\ell(\zb)\big) \enskip .
\end{equation}
The non-negative function $S \mapsto c(S)$ is known as the cohesion function of the product partition probability model.
In a random partition model based on the Dirichlet process, with baseline probability measure $G_0$ and concentration parameter $\alpha$,  one has $c(S)~=~\alpha (|S|-1)!$~\citep{ferguson1973bayesian, sethuraman}. 
	 
Borrowing from \cite{muller2011product},  we define $g(S \given \xb)$ based on an auxiliary probability model $q(\cdot \given \cdot)$, where
\begin{equation} \label{2_2}
    g(S \given \xb) = \int \prod_{i\in S} q(x_i \given \xi)\, \nu(\xi) \,d \xi \enskip .
\end{equation}
Note that the covariates $\xb$ are not random.
The term $\prod_{i\in S}q(x_i \given \xi)$ measures the effect or contribution of the covariates on the prior probability of cluster $S$.
One can choose $q(x_i \given \xi)$ and $\nu(\xi)$ as a conjugate pair to facilitate the analytic evaluation of $g(S \given \xb)$.
For  the cohesion function, we adopt $c(S)=\alpha (|S|-1)!$.

Combining equations~\eqref{2_1} and \eqref{2_2}, we have
\begin{align}\label{eq:prior:z}
    p(\zb \given \xb)\propto 
	\prod^{L(\zb)}_{\ell=1} \bigg[
	\int \!\!\prod_{i\,\in\, S_\ell(\zb)} \!\! q(x_i \given \xi_\ell)\,d\nu(\xi_\ell)
	\bigg] c\big(S_\ell(\zb)\big) \enskip .
\end{align}
In  this model,  $\xi_\ell $ can be considered  the center of the nodal covarites in cluster $\ell$, and $q(x_i \given \xi_\ell)$ a measure of how far the covariates in cluster $S_\ell$ are from its center $\xi_\ell$.
The model then averages over all possible centers $\xi_\ell \sim \nu$. 

The distribution in~\eqref{eq:prior:z} can be written as the marginal of
\begin{align}\label{eq:prior:z:xi}
p(\zb,\xib \given \xb)\propto \prod^{L(\zb)}_{\ell=1}
	\Big[ c\big(S_\ell(\zb)\big)
	\!\!\prod_{i\,\in\, S_\ell(\zb)} q(x_i \mid \xi_\ell)\,\nu(\xi_\ell) \Big] 
	\cdot \!\!
	\prod_{\ell = L(\zb) + 1}^\infty \nu(\xi_\ell) \enskip ,
\end{align}
where $\xib = (\xi_1,\xi_2,\dots)$.
One can use~\eqref{eq:prior:z:xi} to derive a Gibbs sampler for sampling the prior.
To simplify the notation, we let
\begin{align}
	L = L(\zb_{-i}) \quad \text{and} \quad S_\ell = S_\ell(\zb_{-i})
\end{align}
for $\ell \in [L]$ denote the number of clusters of $\zb_{-i} = (z_j, j \neq i)$ and the clusters themselves. Let 
\begin{align}\label{eq:psi:def}
	\psi_k := 
	\begin{cases}
	|S_k| & k \in [L] \\
	\alpha & k = L+1 
	\end{cases} \enskip .
\end{align}
One can show that for $k \in [L+1]$,
\begin{align}
	p(z_i=k, \, \xi_{L+1} \given \zb_{-i},\, \xib_{1:L},\xb) \;\propto\; \nu(\xi_{L+1}) \cdot \psi_k \, q(x_i\given \xi_k) \enskip ,
\end{align}
where $\xib_{1:L} = (\xi_1,\dots,\xi_L)$.
This is equivalent to first drawing $\xi_{L+1} \sim \nu(\cdot)$, and then drawing $z_i$ as follows:
\begin{align}\label{eq:prior:gibbs:2}
    p(z_i=k \given \zb_{-i},\,\xib_{1:L+1},\xb) \;\propto\; \psi_k \, q(x_i\given \xi_k) \enskip .
\end{align}

\begin{example}
    For continuous features, we can take
    \begin{align}\label{eq:Gauss:q:nu}
	    q(x \given \xi)=N(x;\xi,s^2 I) \quad \text{and} \quad \nu = N(0,\tau^2 I) \enskip ,
    \end{align} 
    where $N(x;\xi,s^2 I)$ denotes the density of a normal distribution with mean $\xi$ and covariance matrix $s^2 I$, evaluated at $x$. Then, $k \mapsto q(x_i \given \xi_k)$ in~\eqref{eq:prior:gibbs:2} will be proportional to $\exp(- \norm{x_i-\xi_k}^2/2s^2)$, which shows that if $\xi_\ell$ is the closest to $x_i$ among $\{\xi_k\}$, then the inclusion of the covariate information increases the chance of assigning $z_i$ to cluster~$\ell$.
\end{example}

\begin{example}
    For categorical covariates, one can choose $q(x_i \given \xi_k)$ to be a multinomial distribution and $\nu(\cdot)$ to be a Dirichlet distribution.
    Suppose there are $R$ categorical features and the $r$th feature has $a_r$ categories for $r=1,\ldots, R$.
    Then $x_i=(x_{i1},\ldots, x_{iR})$, where $x_{ir}$ is the $r$-th feature  of node $i$, and $x_{ir} \in \{1,\ldots,a_r\}$.
    Each $\xi_k$ collects parameters of $R$ multinomial vectors, that is, $\xi_k = (\xi_{rk})_{r=1}^{R}$, where the coordinates $\xi_{rk} = (\xi_{rk}^{1},\xi_{rk}^{2},\cdots,\xi_{rk}^{a_r})$ are independent draws from $\text{Dir}(\gamma \bm 1_{a_r})$.
    We have
    \begin{align}\label{eq:Cat:q:nu}
	    q(x_i\mid \xi_k)=\prod_{r=1}^{R}\prod_{c=1}^{a_r}(\xi_{rk}^{c})^{1\{x_{ir}=c\}} \quad \text{and} \quad \nu = \prod_{i=1}^R \Dir(\gamma \bm 1_{a_r}) \enskip .
    \end{align}
    We usually take $\gamma = 1$.
\end{example}

An alternative approach to sample from the prior is to perform Gibbs sampling on the marginalized distribution~\eqref{eq:prior:z}.
This leads to the following updates.
For each $k \in [L+1]$,
\begin{align}\label{eq:prior:gibbs:collapsed}
	p(z_i=k \given \zb_{-i},\,\xb) \;\propto\; \psi_k \,\frac{g(S_k\cup\{i\} \given \xb)}{g(S_k \given \xb)} \enskip ,
\end{align}
where $S_{L+1} = \emptyset$ and $g(\emptyset \given \xb) =1$. This approach is, in particular, useful when $q(\cdot \given \cdot)$ and $\nu(\cdot)$ are conjugate so that $g(S \given \xb)$ is easy to compute.
	
With the priors thus defined, the network is assumed to follow a SBM, that is, 
\begin{align}\label{eq:SBM}
    p(A \given \etab, \zb)=\prod_{1\leq i<j\leq n} \eta_{z_i,z_j}^{A_{ij}}(1-\eta_{z_i,z_j})^{1-A_{ij}} \enskip ,
\end{align}
where $\etab = (\eta_{k,\ell})$ is the connectivity matrix of SBM, with $\eta_{k,\ell}$ representing the link probability between nodes in clusters $k$ and $\ell$. 

It is possible to obtain closed forms for the full conditional distributions of the unknown model parameters $\boldsymbol \eta, \zb$, and $\xib$ with appropriate choices of $q(\cdot)$ and $\nu(\cdot)$, as demonstrated in Section~\ref{sec:gibbs}. We note that since the prior random partition model~\eqref{2_1} puts mass on all potential partitions of the $n$ nodes, the posterior distribution $p(\zb \given A, \xb)$ also puts mass on all such partitions; however, the posterior will be concentrated around certain partition(s), hence a posteriori, there is a most likely value of $L(\zb)$, the number of communities in $\zb$. That is how the model learns the number of communities. 

\subsection{Comparison with literature}\label{sec:comparison}

There are several works in the literature similar to our model that also use a Bayesian nonparametric approach for tasks related to node clustering.

A pioneering work in the area is that of \cite{kemp}.
Motivated by the complex system of relations underlying semantic knowledge, \cite{kemp} propose the infinite relational model (IRM) for discovering and clustering underlying structure in relational data sets.
In this framework, the observed data are assumed from $n$ types (people, demographic features, answers to a personality test, etc.) and $m$ relationships (person $i$ likes person $j$, feature $x$ causes answer $y$, etc).
The motivating example given in \cite{kemp} is that of clustering people, represented by set $T^1$, based on social predicates, represented by set $T^2$. The observed data is the tensor $T^1 \times T^1 \times T^2 \mapsto \{0,1\}$ whose $(i,j,p)$ entry determines whether persons $i$ and $j$ have social predicate type $p$.
The idea is to simultaneously cluster $T^1$ and $T^2$ so that the tensor is roughly constant within the resulting clusters.
In modern language, the model proposed by~\cite{kemp} is the so-called \emph{tensor SBM} \citep{kim2017community, wang2019multiway, lei2020consistent} but with a CRP prior on the labels of each dimension to allow for infinite clusters a priori.
IRM was later explicitly adopted in \cite{BCD2012} for community detection in network data, as opposed to relational data.
IRM is quite flexible and can, for example, be used to incorporate an attribute or feature taking finite values, by taking $T^2$ above to be the levels of that attribute. However, this attribute should be interpreted as an edge feature, and moreover, IRM needs to have observations on the connectivity of persons $(i,j)$ for all possible levels of this attribute. Adding each feature then requires increasing the dimension of the tensor by one, and demanding lots of observations which are not available in practice in network problems.


More recently, nonparametric Bayesian network models that accommodate nodal covariates have been considered, but mostly with the mixed membership SBM (MMSBM) framework of \cite{airoldi2008mixed}. For example, \cite{meta}
introduced the nonparametric metadata dependent relational (NMDR) model, that essentially couples the MMSBM likelihood with node-covariate-dependent prior on cluster labels. More specifically, they assume latent community vectors $\eta_k$ that interact with node features $\phi_i$ to produce a score $v_{ki}$ that determines how likely node $i$ belongs to community $k$, a priori. They assume that $v_{ki}$ are normal with mean $\langle \eta_k, \phi_i \rangle$ and then translate these real-valued affinities to probabilities $\pi_{ki}$ via a logistic-stick breaking process \citep{logistic-stick}. The $\pi_i = (\pi_{ki})$ then determine the edge probabilities via $\mathbb E [A_{ij} \given \pi_i, \pi_j] = \pi_i^T W \pi_j$ where $W$ is the connectivity matrix.


Along the same lines, \cite{pmlr-v70-zhao17a} extends the edge partition model (EPM) of~\cite{zhou2015infinite} to incorporate binary node features. The EPM has similarities to MMSBM
with novel uses of a Bernoulli-Poisson likelihood coupled with a nonparametric partition model. More specifically, the latent Poisson component $X = (X_{ij})$ still follows a MMSBM decomposition with $\mathbb E[X_{ij} \given \phi_i, \phi_j] = \phi_i^T \Lambda \phi_j$ where $\phi_i$ are the soft community assignments and $\Lambda$ the connectivity matrix.  Similar to~\cite{meta}, the nodal covariate information is incorporated in constructing a prior on $\phi_i = (\phi_{ik})$. The prior assumes $\phi_i$ to be drawn from a Gamma distribution with mean $\mathbb E[\phi_{ik}] = c_i b_k \prod_{\ell=1}^L h_{\ell k}^{f_{i \ell}}$ where $f_{i \ell}$ is the $\ell$th binary feature of node $i$. Note that by introducing $\eta_k := (\log h_{\ell k})$ and $f_i = (f_{i\ell})$, one can write $\mathbb E[\phi_{ik}] = c_i b_k \exp(\langle \eta_k, f_i\rangle)$, showing that essentially the same inner product interaction of feature and latent community vectors as in~\cite{meta} is used by \cite{pmlr-v70-zhao17a}.

As in \cite{meta} and \cite{pmlr-v70-zhao17a}, 
our model also incorporates node features into the partition prior; however, we are modeling hard community assignments rather than soft assignment vectors, making the problem somewhat more challenging. More importantly, our approach allows for a more general dependence of the partition on the features via a kernel $q(x_i \given \xi)$, compared to the simple inner product approach used in both \cite{meta} and \cite{pmlr-v70-zhao17a}.
Our approach is  not limited to binary features and by incorporating more complexity into $q(x_i \given \xi)$, we can potentially model more complex feature/community interactions in the prior.


The last closely related work that we discuss is that of \cite{newman2016structure}. They consider node features (metadata) that take values in a finite discrete set (say $\mathbb X$). Similar to our work, the node metadata is used to influence the prior on the community assignements. In our notation, they assume $p(z_i \given x_i) = \gamma_{z_i, x_i}$ where $\gamma~=~(\gamma_{k,x}) \in~[0,1]^{K \times |\mathbb X|}$ is a parameter matrix to be estimated form the data. Compared to the inner product model of~\cite{meta} and~\cite{pmlr-v70-zhao17a}, this approach gives a more flexible model for the interaction of the communities and  features. The drawback is that it is limited to discrete features and if $|\mathbb X|$ is large, there is potential for over-fitting without further regularization of the $\gamma$ matrix. \cite{newman2016structure} use an EM algorithm to estimate the parameters, and they assume the number of communities to be known.


\subsection{Gibbs sampler}
\label{sec:gibbs}

We now derive a Gibbs sampler to sample from the complete posterior distribution of $(\zb,\xib,\bm \eta)$ given $A$ and $\xb$.
The main challenge is deriving the updates for $\zb$.

We sample from $\zb$, $\xib$, and $\boldsymbol \eta$ through their full conditional distributions, which are given bellow, until reaching convergence, and then obtain a sample of adequate size of the posterior distribution for inference.

\subsubsection{Initialization}

We initialize the labels by drawing from a Chinese Restaurant Process (CRP),
\begin{align*}
	\zb \sim \CRP(\alpha) \enskip .
\end{align*}
A CRP can be seen as a special case of our prior without any covariates.
This follows from~\eqref{eq:prior:gibbs:collapsed} by setting $g(S \given \bm x) =1$.
Once $\zb$ is initialized, all the other parameters can be initialized by the Gibbs updates derived below.

Note that initializing the chain by sampling $\zb$ from a CRP provides a random start without having to specify the number of the communities $K$.
In many algorithms, spectral clustering is often used to initialize $\zb$.
For a Bayesian model, this is not a natural choice.
Moreover, it requires the knowledge or an estimate of $K$.

\subsubsection{Sampling $\zb$} 
Let $b(x;a,b) = x^{a-1} (1-x)^{b-1}$ and for simplicity, define
\begin{align} \label{eq:ct:def}
	\ct_\ell(S) := c(S) \prod_{j \in S} q(x_j \given \xi_\ell) \,\nu(\xi_\ell) \enskip .
\end{align}
Note that $\ct_\ell(S)$ implicitly depends on $\xi_\ell$.
We have
\begin{align*}
	&p(A, \zb, \xib, \boldsymbol \eta \mid \xb) = p(A \mid \boldsymbol \eta, \zb) \cdot p(\zb, \xib \mid \xb) \cdot p(\boldsymbol \eta) \\
	&\qquad \propto \prod_{1 \le i < j \le n} 
	\eta_{z_i z_j}^{A_{ij}} (1- \eta_{z_i z_j})^{1-A_{ij}}
	\prod_{\ell = 1}^\infty \ct_\ell(S'_\ell)
	\prod_{1 \le m \le \ell \le \infty} 
	b(\eta_{\ell m}; \beta, \beta) \enskip .
\end{align*}

Here, $S'_\ell = \{i \in [n]:\; z_i = \ell\}$ is the $\ell$th community of $\zb$.
We assume that $\zb$ has $L'$ communities $S'_1,S'_2,\dots,S'_{L'}$ and let $S'_{L'+1} = S'_{L'+2} = \cdots = \emptyset$.
The convention is that $c(\emptyset) = 1$ while $c(S) = \alpha \Gamma(|S|)$ when $S$ is nonempty.
Similarly, $\prod_{j \in \emptyset} (\,\cdots) = 1$.
In the above, we assume that $\xib = (\xi_1,\xi_2,\dots)$ collects all possible $\xi_\ell$ and similarly for $\etab = (\eta_{\ell m}: \ell, m \in \mathbb{N})$.

Fix $i$ and let $S_\ell = \{j \in [n]\setminus i: z_j = \ell\}$ be the $\ell$th community of $\zb_{-i}$.
We assume that $\zb_{-i}$ has $L$ communities $S_1,S_2,\dots,S_L$ and by convention, let $S_{L+1} = S_{L+2} = \dots = \emptyset$. 
To get the communities of $\zb$ from $\zb_{-i}$, we either update $S_k$ to $S'_{k} = S_k \cup \{i\}$ for some $k \in [L]$, or update $S_{L+1}$ to $S'_{L+1} = S_{L+1} \cup \{i\} = \{i\}$, generating a new community.
With the convention we use, we can compactly write both cases as $S'_k = S_k \cup \{i\}$ for all $k \in [L+1]$.

For any $k \in [L+1]$, we obtain
\begin{align}\label{eq:zi:conditional:expanded}
	\begin{split}
		&p\big(z_i = k, \,\xib_{L+1}, \,\bm \eta_{L+1,[L]} \given \zb_{-i},\, A, \,\xib_{[L]}, \,\bm \eta_{[L],[L]}, \, \xb\big) 
	\propto \\
	&\quad \prod_{j: j \neq i} 
		\eta_{k, z_j}^{A_{ij}} (1- \eta_{k, z_j})^{1-A_{ij}}
	 \cdot \frac{\ct_k(S_k \cup \{i\})}{\ct_k(S_k)}\prod_{\ell = 1}^{L+1} \ct_\ell(S_\ell) \prod_{1 \le m \le \ell \le L+1} 
	\!\!\! b(\eta_{\ell m}; \beta, \beta) \enskip .
	\end{split}
\end{align}
Letting
\begin{align}\label{eq:O:n:def}
	O_{i\ell} = \sum_{j: j \neq i}  A_{ij} 1\{z_j = \ell\}, \quad 
	n_\ell = \sum_{j: j\neq i}1\{z_j = \ell\} \enskip ,
\end{align}
we have $\prod_{j: j \neq i} 
\eta_{k, z_j}^{A_{ij}} (1- \eta_{k, z_j})^{1-A_{ij}}
 = \prod_{\ell = 1}^L \eta_{k\ell}^{O_{i\ell}} (1-\eta_{k\ell})^{n_{\ell} - O_{i\ell}}$.

Noting that $\prod_{\ell = 1}^{L} \ct_\ell(S_\ell)$ is a constant in~\eqref{eq:zi:conditional:expanded}, and similarly for any term in 
\[\prod_{1 \le m \le \ell \le L+1} b(\pi_{\ell m}; \beta, \beta)\]
that does not have an index equal to $L+1$, we obtain
\begin{align}
	\begin{split}
	&p\big(z_i = k, \,\xib_{L+1}, \,\bm \eta_{L+1,[L]} \given \zb_{-i},\, A, \,\xib_{[L]}, \,\bm \eta_{[L],[L]}, \, \xb\big)
	\propto \\
	&\quad	\prod_{\ell = 1}^L \eta_{k\ell}^{O_{i\ell} } (1-\eta_{k\ell})^{n_{\ell} - O_{i\ell}}\cdot \frac{\ct_k(S_k \cup \{i\})}{\ct_k(S_k)} \ct_{L+1}(S_{L+1})  
	\prod_{m=1}^{L} b(\eta_{L+1, m}; \beta, \beta) \enskip .
	\end{split}
\end{align}
The product over $m$ runs up to $L$ since only $\eta_{L+1,[L]}$ is a variable while $\eta_{L+1,L+1}$ is a constant. This is because $z_i$ can take the new value $L+1$ but $z_j$ with $j \neq i$ takes values in $[L]$, hence we do not need to sample $\eta_{L+1,L+1}$ at this stage.

Since $S_{L+1} = \emptyset$, we have
\begin{align*}
	\ct_{L+1}(S_{L+1}) &= \nu(\xi_{L+1}) \enskip , \\ 
	\ct_{L+1}(S_{L+1} \cup \{i\})  &= \alpha \Gamma(1) \,q(x_i \given \xi_{L+1}) \,\nu (\xi_{L+1}) \enskip .
\end{align*}
Hence for $k \in [L+1]$,
\begin{align*}
	\frac{\ct_k(S_k \cup \{i\})}{\ct_k(S_k)} \ct_{L+1}(S_{L+1})  
	&= \nu (\xi_{L+1}) \, \psi_k \, q(x_i \mid \xi_k) \enskip ,
\end{align*}
where $\psi_k$ is defined in~\eqref{eq:psi:def}.
Thus, we can compactly write
\begin{align}
    \begin{split}
        &	p\big(z_i = k, \,\xib_{L+1}, \,\bm \eta_{L+1,[L]} \given \zb_{-i},\, A, \,\xib_{[L]}, \,\bm \eta_{[L],[L]}, \, \xb\big) \propto \\
        &\quad \nu(\xi_{L+1}) \prod_{m=1}^{L}
        b(\eta_{L+1, m}; \beta, \beta)
	    \cdot \psi_k q(x_i \given \xi_k)\prod_{\ell = 1}^L \eta_{k\ell}^{O_{i\ell}} (1-\eta_{k\ell})^{n_{\ell} - O_{i\ell}} \enskip .
    \end{split}
\end{align}
This is equivalent to the following.
First draw $\xi_{L+1} \sim \nu$ and $\eta_{L+1, m} \sim \text{Beta}(\beta,\beta)$ for $m \in [L]$, all independently.
Then draw $z_i$ from 
\begin{align}\label{eq:zi:update}
\begin{split}
	&p\big( z_i = k \given \zb_{-i}, \,A,\, \xib_{[L+1]}, \bm \eta_{[L+1],[L]}, \xb\big) \;\propto\;  \\
	&\qquad \qquad  \psi_k \, q(x_i \mid \xi_k)	\prod_{\ell = 1}^L \eta_{k\ell}^{O_{i\ell}} (1-\eta_{k\ell})^{n_{\ell} - O_{i\ell}}, \quad k \in [L+1] \enskip .
\end{split}	
\end{align}
For continuous features, we use~\eqref{eq:Gauss:q:nu} for $q(\cdot \given \cdot)$ and $\nu$, and for categorical variables we use~\eqref{eq:Cat:q:nu} in the above updates. 

\begin{remark}
Note that update~\eqref{eq:zi:update} is where a potentially new community (labeled $L+1$) is created. This happens if the following conditions are met: (a) when sampling $z_i$ according to~\eqref{eq:zi:update}, we happen to pick $z_i = L+1$, (b) $L' = L$, that is, the current number of communities is $L$, and (c) currently $z_i$ is not assigned to a singlton community. In such a case, the number of communities will increase from $L'=L$ to $L+1$. On the other hand, update~\eqref{eq:zi:update} can also annihilate a community if the following holds: (a) When sampling $z_i$, we pick $z_i \in [L]$ and (b) $L' = L+1$ which means that $z_i$ is currently assigned to a singleton community. In this case, the new number of communities will go from $L'$ to $L'-1 = L$.
\end{remark}

\subsubsection{Sampling $\xib$}

We have
\begin{align*}
	p(\xib \given A, \zb, \xb, \bm \eta) &= p(\xib \mid \zb, \xb)  \propto \prod_{\ell = 1}^{L'} H_{ S'_\ell}(\xi_\ell) \enskip ,
\end{align*}
where $H_S$ is the distribution with density 
\begin{align}
	H_S(\xi) &\propto  \prod_{i \in S} q(x_i \given \xi) \,\nu(\xi) \enskip .
\end{align}  
That is, $\xi_\ell$ are independent draws from $H_{S'_\ell}$.
We recall that $S'_\ell = \{i \in [n]:\; z_i = \ell\}$.

The details of sampling $\xib$ are slightly different given different choices of $q(\cdot)$ and $\nu(\cdot)$ depending on whether continuous or categorical features are available. 
For continuous features with the Gaussian choice~\eqref{eq:Gauss:q:nu}, $H_S(\xi) \propto \prod_{i\in S}N(x_i;\xi,s^2 I)\cdot N(\xi;0,\tau^2 I)$ which gives
\begin{align*}
	H_S =  N\left(\frac{\tau^2\sum_{i\in S}x_i}{|S| \tau^2+s^2},\frac{s^2\tau^2}{|S|\tau^2+s^2} I \right) \enskip .
\end{align*}
For the categorical features with the choice~\eqref{eq:Cat:q:nu}, we have
\begin{align*}
	H_S(\xi) \propto \prod_{i \in S} \prod_{r=1}^R \prod_{c=1}^{a_r} (\xi_r^c)^{1\{x_{ir} = c\}} \cdot \prod_{r=1}^R \prod_{c=1}^{a_r} (\xi_r^c)^{\gamma - 1}
	 =  \prod_{r=1}^R \prod_{c=1}^{a_r}  (\xi_r^c)^{\alpha_r^c(S) - 1} \enskip ,
\end{align*}
where $\alpha_r^c(S) := \gamma + \sum_{i \in S} 1\{x_{ir} = c\}$.
That is, $H_S$ is the product of Dirichlet distributions,
\begin{align*}
	H_S = \prod_{r=1}^R \Dir(\alpha_r^1(S),\dots, \alpha_r^{a_r}(S)) \enskip .
\end{align*}

\subsubsection{Sampling $\etab$}

Let us define the index sets
\begin{align}\label{eq:Gamma:index}
	\Gamma_{k\ell} = 
	\begin{cases}
		\{(i,j) : 1 \le i < j \le n\} & k = \ell \\
		\{(i,j) : 1 \le i \neq j \le n\} & k \neq \ell
	\end{cases} \enskip ,
\end{align}
and block counts
\begin{align}\label{eq:M:N:counts}
	M_{k\ell} = \sum_{(i,j) \in \Gamma_{k\ell}} 
	A_{ij} 1\{z_i = k, z_j = \ell\},
	\quad 
	N_{k \ell} =  \sum_{(i,j) \in \Gamma_{k\ell}} 1\{z_i = k, z_j = \ell\} \enskip .
\end{align}
Then, we have
\begin{align*}
    p(\etab \mid A, \zb, \xb, \xib) = p(\etab \mid A, \zb)  \propto \prod_{k \le \ell} \eta_{k\ell}^{M_{k\ell}+\beta-1}(1-\eta_{k\ell})^{N_{k\ell} - M_{k\ell} + \beta-1} \enskip .
\end{align*}
Thus, $\eta_{k\ell}$ are independent draws from $\Beta(M_{k\ell} +\beta, N_{k\ell} - M_{k\ell} + \beta)$.

\begin{remark}[Directed networks]
Although our current MCMC algorithm is only for undirected networks, our prior can be used for Bayesian community detection in directed networks with covariates.
One can simply replace the undirected SBM likelihood in~\eqref{eq:SBM} with a directed SBM, by replacing $i < j$ with $i\neq j$ and removing the symmetry constraint on $\etab$. The resulting sampler will be almost identical, except for minor modifications to the counts~\eqref{eq:O:n:def} and~\eqref{eq:M:N:counts} to account for the extra edge information.
\end{remark}

\section{Simulation study} \label{sec:simu}
	
In this section, we carry out multiple simulation studies in which we compare our methods, which we refer to as BCDC (Bayesian community detection for networks with covariates), with i) the covariate-assisted spectral clustering (CASC) algorithm \citep{binkiewicz2017covariate}, which uses both the network and the covariates information in a spectral clustering algorithm, ii) the covariated-assisted clustering on ratios of singular vectors (CASCORE) algorithm \citep{hu2022covariate}, which modfies CASC for degree heterogeneity, iii) $k$-means algorithms ($k$-means) applied only to the covariates, iv) spectral-clustering (SC) of the adjacency matrix, and v) a Bayesian SBM (BSBM), which is essentially a special case of our model with $g(S\mid \boldsymbol x)=1$, therefore utilizing only the network information. 
For CASC, the core idea is to first construct a new Laplacian matrix $L_x=L+\tau XX^T$, where $L$ is the Laplacian matrix for the network and $X$ denotes the $n$ by $p$ feature matrix, and then apply the standard spectral clustering algorithm on $L_x$.
Throughout, we select $\tau$ based on the automated procedure given in \citep[Section 2.3]{binkiewicz2017covariate}. 

We consider simulation designs with (a) continuous features, (b) discrete or categorical features, (c) a mix of continuous and discrete covariates, (d) high-dimensional features, and (e) homophily effects with networks simulated from stochastic block models with varying connectivity patterns and sparsity levels.
The performance of the estimated communities is measured by normalized mutual information (NMI), a measure ranging from 0 (random guessing) to 1 (perfect agreement).
NMI is a  measure of similarity of two partitions, and is widely used in the community detection literature. 
It allows comparisons of two partitions with different number of clusters while accounting for the issue of label invariance.

To define NMI, consider two partitions (labelings) on a set of objects and let $(X,Y)$ be the two labels of a randomly drawn object. The joint probability distribution of $(X,Y)$ is the normalized confusion matrix between the two partitions. We can define the mutual information $I(X,Y)$ and joint and marginal entropies---$H(X,Y)$, $H(X)$ and $H(Y)$---based on the aforementioned joint distribution, using standard definitions. It is common to define NMI as $I(X,Y) / H(X,Y)$. However, there are other variants and to be consistent with prior work, in particular~\cite{sarkar}, we will use the variant implemented in the \texttt{R} package \texttt{NMI}, namely, $2 I(X,Y) / (H(X) + H(Y))$, which is also referred to as \emph{symmetric uncertainty}~(\citet[p. 634]{teukolsky1992numerical}; \cite{hall1998correlation}).

Overall, the simulation studies show that our method consistently outperforms the competitors and demonstrates the gain of our model by utilizing both the network and nodal information for detecting the community structures.
The simulations were performed on a high-computing cluster.
An R package for our samplers, as well as the code for these experiments, is available at the GitHub repository \href{https://github.com/aaamini/bcdc}{aaamini/bcdc}~\citep{Shen_BCDC_model_for_2022}.
We ran our code on a high-performance cluster with an Intel(R) Xeon(R) CPU E5-2680 v4 @ 2.40GHz with 28 cores and 256 GB RAM.

\subsection{Continuous covariates}
	
We first consider simulated networks with continuous covariates, and in particular, the Gaussian setting~\eqref{eq:Gauss:q:nu}.
We generate networks from an SBM having connectivity matrix $\eta = (\eta_{k\ell}) \in [0,1]^{K \times K}$ with
\begin{align}\label{eq:pp:eta}
	\eta_{k\ell} = 
	\begin{cases}
		p & k = \ell \\
		rp & k \neq \ell
	\end{cases} \enskip .
\end{align}
The parameter $r \in [0,1]$ controls the magnitude of disparity between the within and between connectivites and is a measure of network information for the community structure.
In our simulations, we set $p = 0.1$ and vary $r$.
We consider $n=150$ nodes with $K=2$ communities of 100 and 50 nodes, respectively.
	
For each node, we generate $d=2$ features, with one \emph{signal} feature related to the community structure and one \emph{noise} feature whose distribution is the same for all nodes.
Letting $x_i \in \reals^2$ be the feature vector for node $i$ and $e_1 = (1,0)$, we take
\begin{align*}
	x_i \given z_i \sim N\bigl( 
	\mu \sigma_{z_i} e_1,  I_2	\bigr) \enskip ,
\end{align*}
where $\sigma_1 = +1$, $\sigma_2 = -1$ and $z_i \in \{1,2\}$ is the community label of node~$i$. Here $\mu \in [0, \infty)$ is proportional to the signal-to-noise ratio of the covariate information.

\begin{figure}[t!]
	\includegraphics[width = 0.49\textwidth]{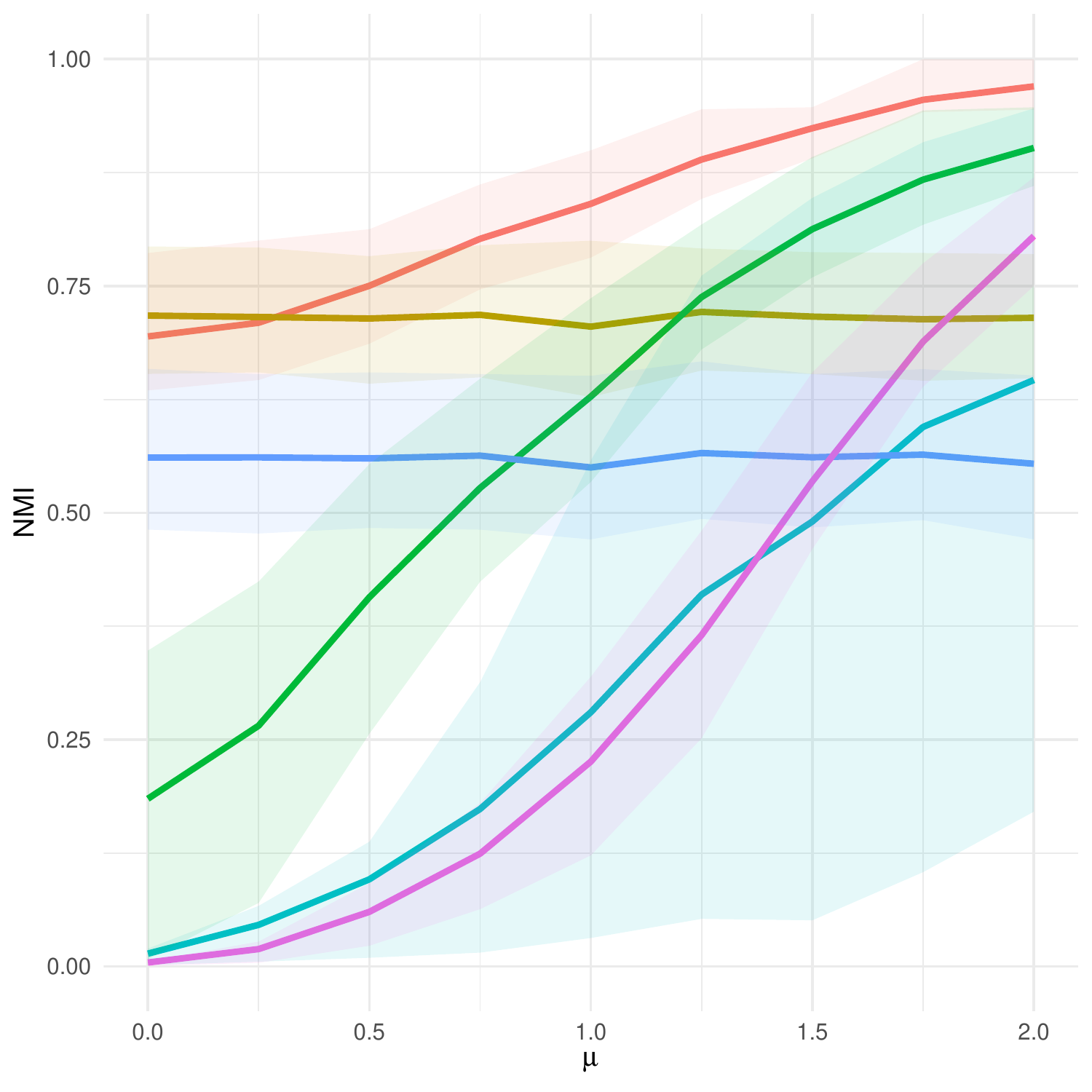}
	\includegraphics[width = 0.49\textwidth]{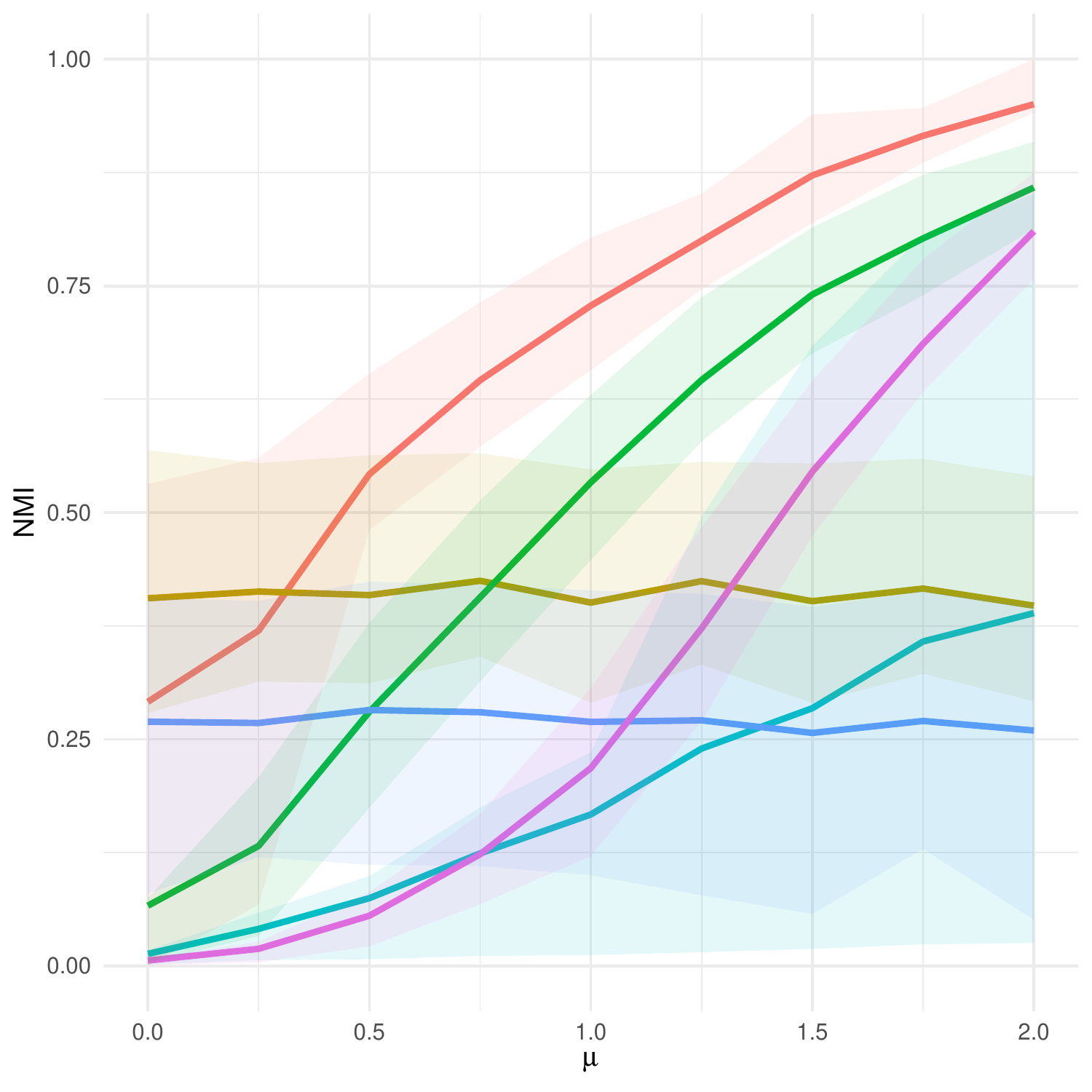}
	\includegraphics[width = 0.49\textwidth]{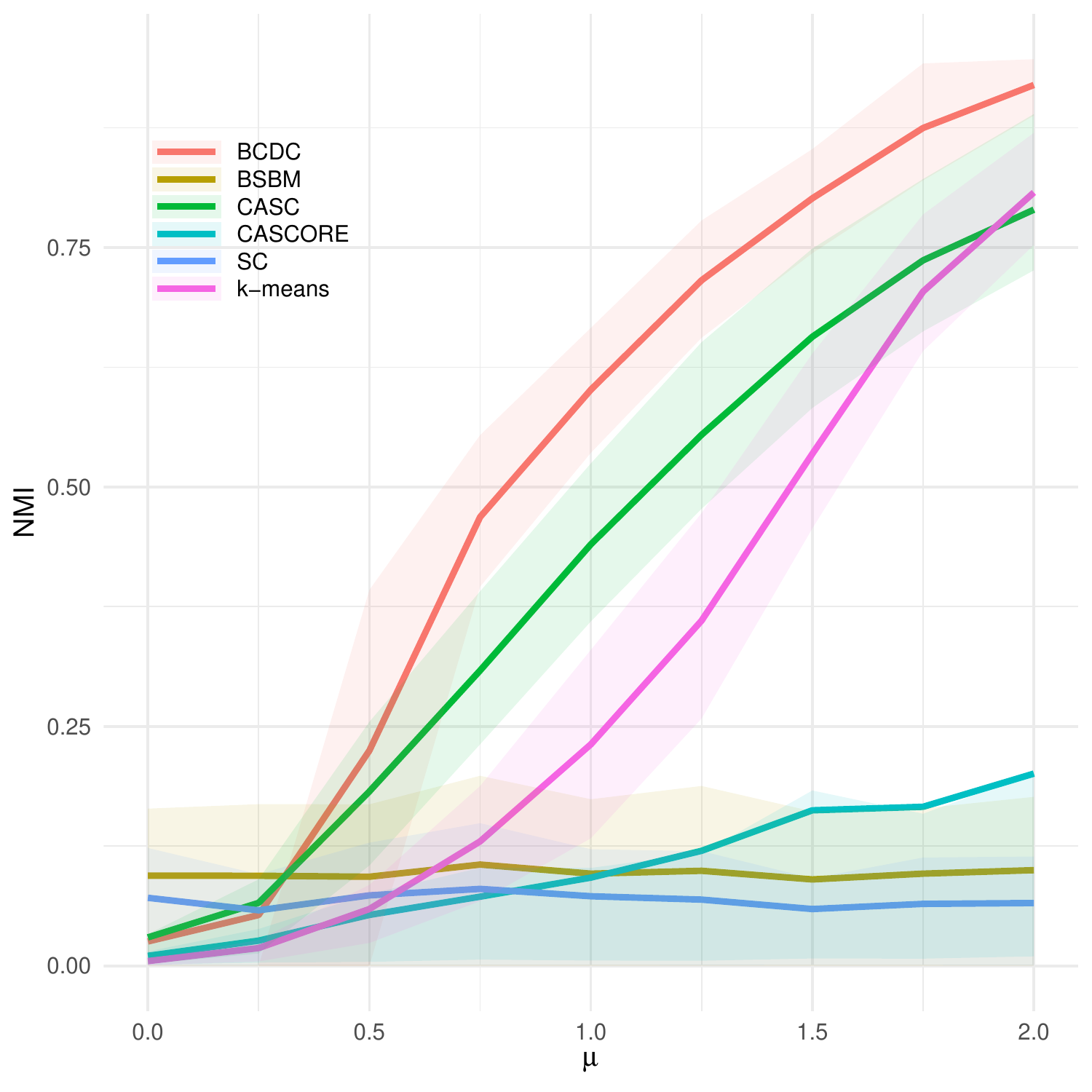}
	\includegraphics[width = 0.49\textwidth]{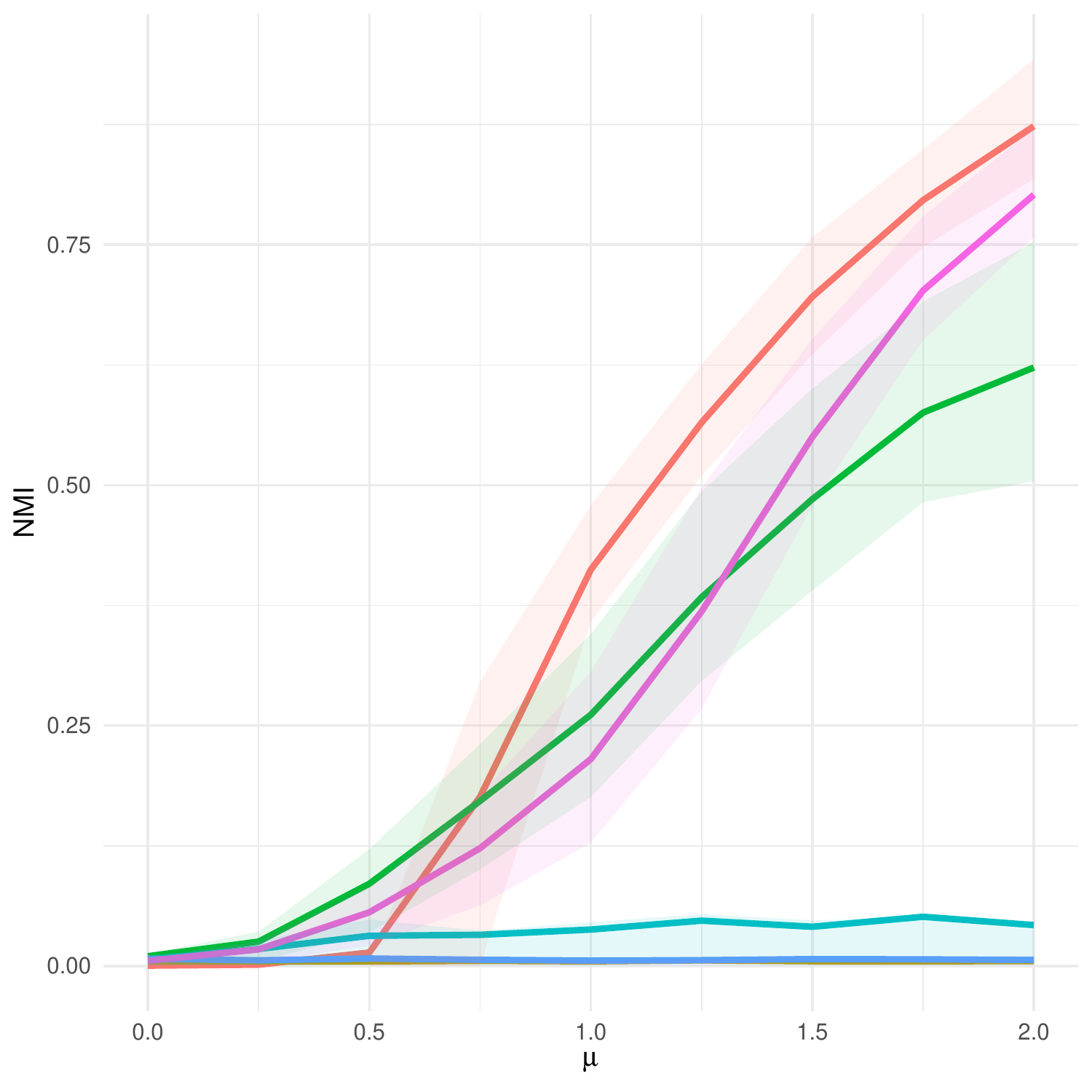}
	\caption{NMI results for five different methods on a 2-block SBM with continuous data. In all cases, $p = 0.1$, and we vary the network and covariate signal-to-noise ratios, $r$ and $\mu$, respectively.}
	\label{fig:boxplot}
\end{figure}

Figure \ref{fig:boxplot} shows the mean NMI with a 50\% quantile band from BCDC and competing methods, averaged over 500 replications, under different settings of $r$ and $\mu$.
For BCDC, we have used parameters $\alpha = 10$, $\beta = 1$ and $\tau = s =1$, and ran the sampler for 1000 iterations.
Note that in our comparison, all of the competing methods were given an additional advantage by assuming the knowledge of the true number of communities.
However, our method (red) consistently outperforms these other methods.
An interesting notable case is that of high network information ($r = 0.3$) and pure noise covariate information ($\mu = 0$).
In this case, BCDC performs as well as BSBM which only operates on network information, while CASC performs much worse being misled by pure noise covariates.

\subsection{Categorical covariates}
	
We next consider a simulation study for networks with categorical covariates.
For each node $i$, we again generate $d=2$ features with one \emph{signal} feature related to the community structure and one \emph{noise} feature whose distribution is the same for all nodes.
We consider two designs:
\begin{enumerate}[(\arabic*), wide]
	\item We consider networks with $n = 150$ nodes and $K = 3$ equally-sized communities.
	The signal features are taken to be the true community labels and the noise features are uniformly distributed on $\{1,2,3\}$.
	\item We consider networks with $n = 150$ nodes and $K = 2$ communities of 100 and 50 nodes.
	We create two 4-category features.
	Let $x_i = (x_{i1}, x_{i2}) \in \{1,2,3,4\}^2$ be the feature vector for node~$i$.
	We use the following generative model
	\begin{align*}
		\theta_1, \theta_2 &\sim \Dir(\bm 1_4) \enskip , \\
		x_{i1} \given z_i &\sim \theta_{z_i}, \quad x_{i2} \given z_i \sim \bm 1_{4}/4 \enskip ,
	\end{align*}
	where, for example, $x_{i1} \sim \theta_{z_i}$ means that $x_{i1}$ is a categorical variable with probability vector $\theta_{z_i}$. 
\end{enumerate}
In both cases, we use an SBM with connectivity~\eqref{eq:pp:eta}, setting $p = 0.1$ and varying $r$ from 0.1 to 0.8.
For the parameters of our model, we again set $\alpha=10$, and ran the chain for 1,500 iterations.
As above, the results are given over 500 replicates.
	
\begin{figure}
	\includegraphics[width=.49\textwidth]{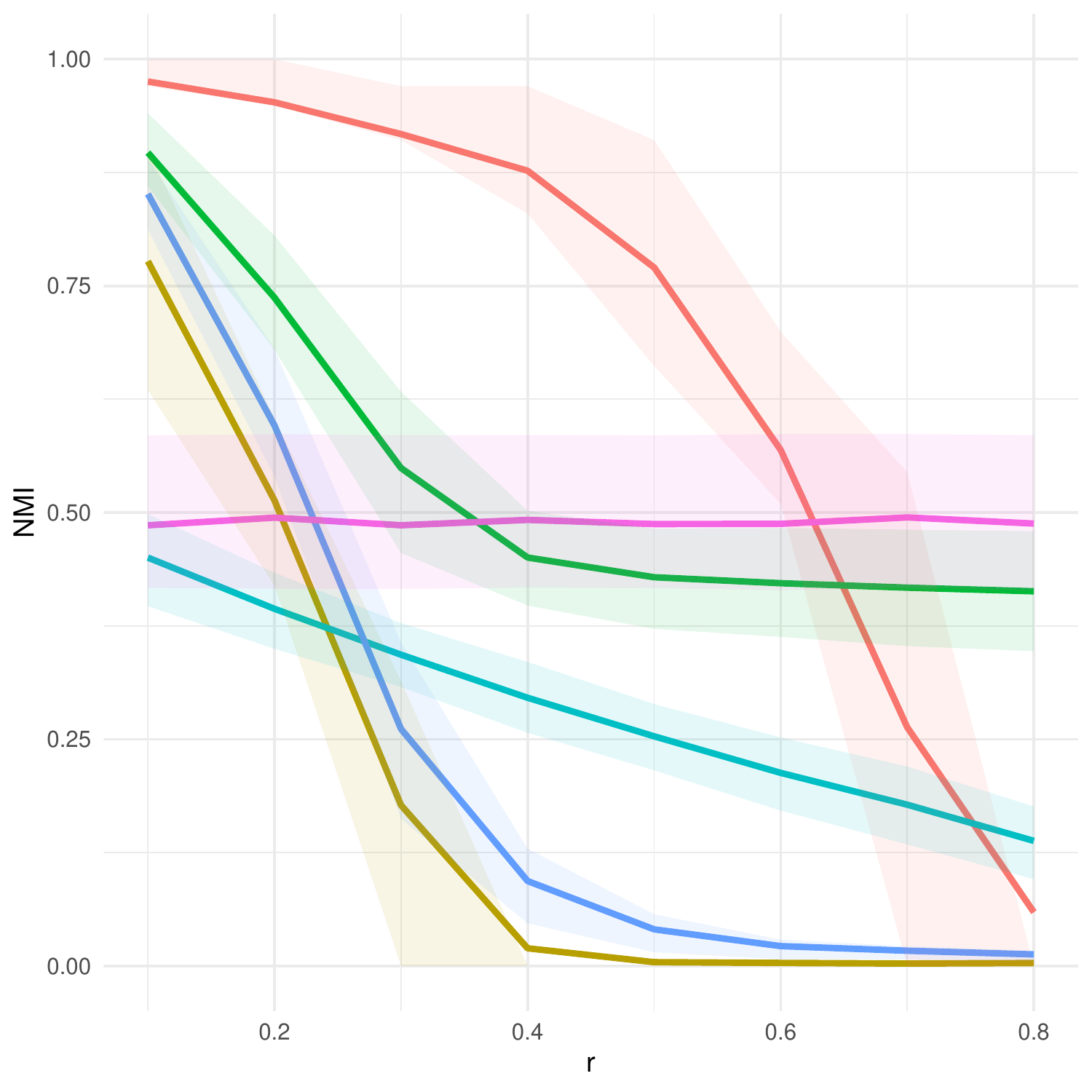}
	\includegraphics[width=.49\textwidth]{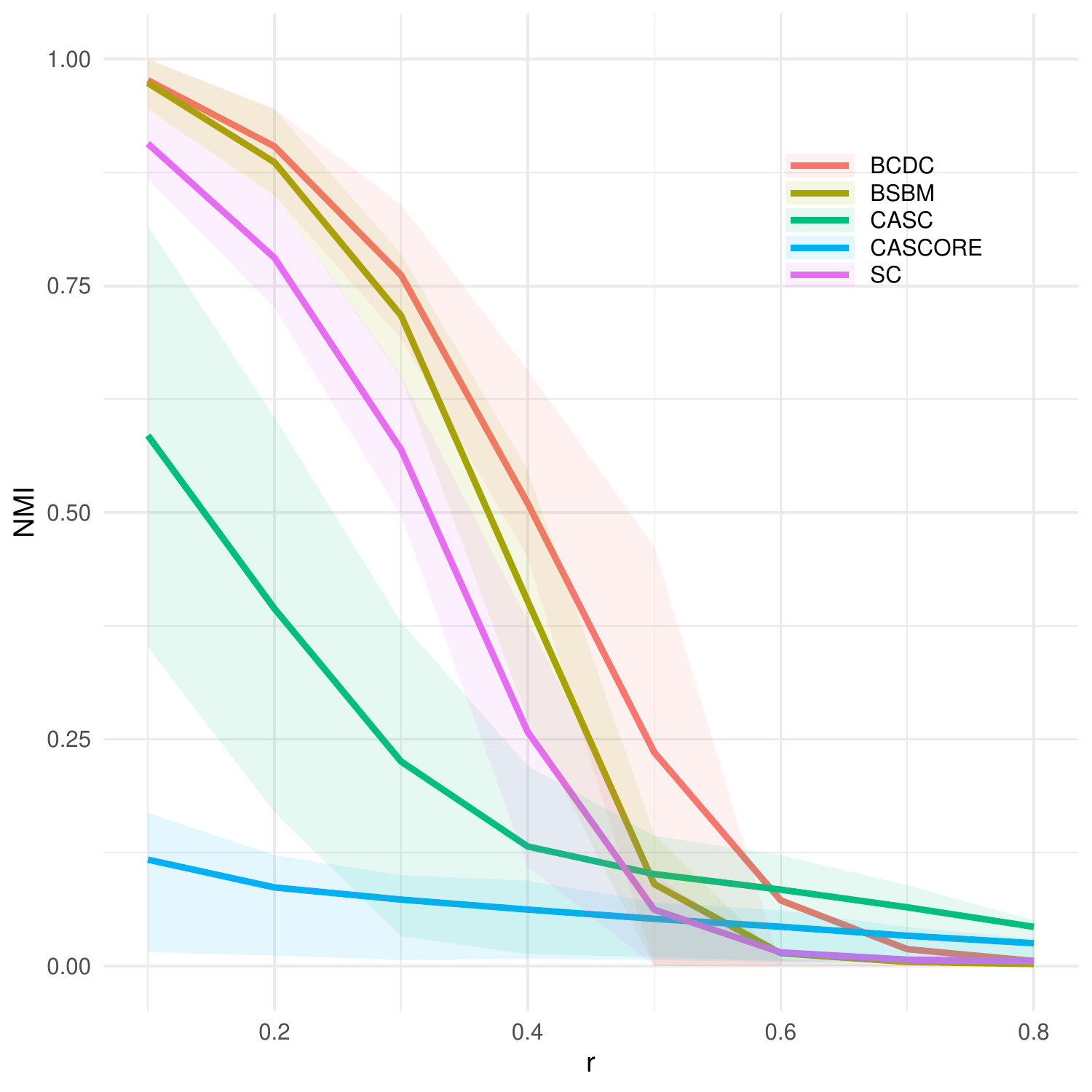}
    \caption{NMI results for five different methods on a 3-block (left) and 2-block (right) SBM with categorical data.}
    \label{fig:boxplot_cat}
\end{figure}

Figure \ref{fig:boxplot_cat} shows NMI as a function of $r$ (the network information measure) under the two covariate designs.
Once again, all methods except BCDC were given the true number of communities.
The NMI values obtained under the proposed model are generally higher than those of the other models with a slightly larger variance, which is likely due to the additional uncertainty in estimating the number of the communities.

\subsection{Mix of continuous and discrete covariates}

Here, we perform a simulation for larger networks with more communities and a mix of continuous and categorical variables.
We let the number of nodes $n$ vary from 300 to 1000 and set $K = n/50$ communities.
The features are chosen so that neither perfectly separates the clusters, but both are informative.
In particular, we take
\begin{align*}
	x_{1i} \given z_i \sim N\bigl( 
	2(z_i \bmod 2) - 1,  1	\bigr) \enskip ,
\end{align*}
i.e. $x_{1i} \sim N(1, 1)$ for $z_i \in \{2, 4, \ldots\}$ and $x_{1i} \sim N(-1, 1)$ for $z_i \in \{1, 3, \ldots\}$.
We also take $x_{2i} = 1$ for $z_i \in \{1, 2, \ldots K/2\}$ and $x_{2i} = 2$ for $z_i \in \{K/2+1, \ldots, K\}$.
As before, we use an SBM with connectivity \eqref{eq:pp:eta}, setting $p = 0.3$ and $r = 0.35$.

\begin{figure}[t!]
	\includegraphics[width=.49\textwidth]{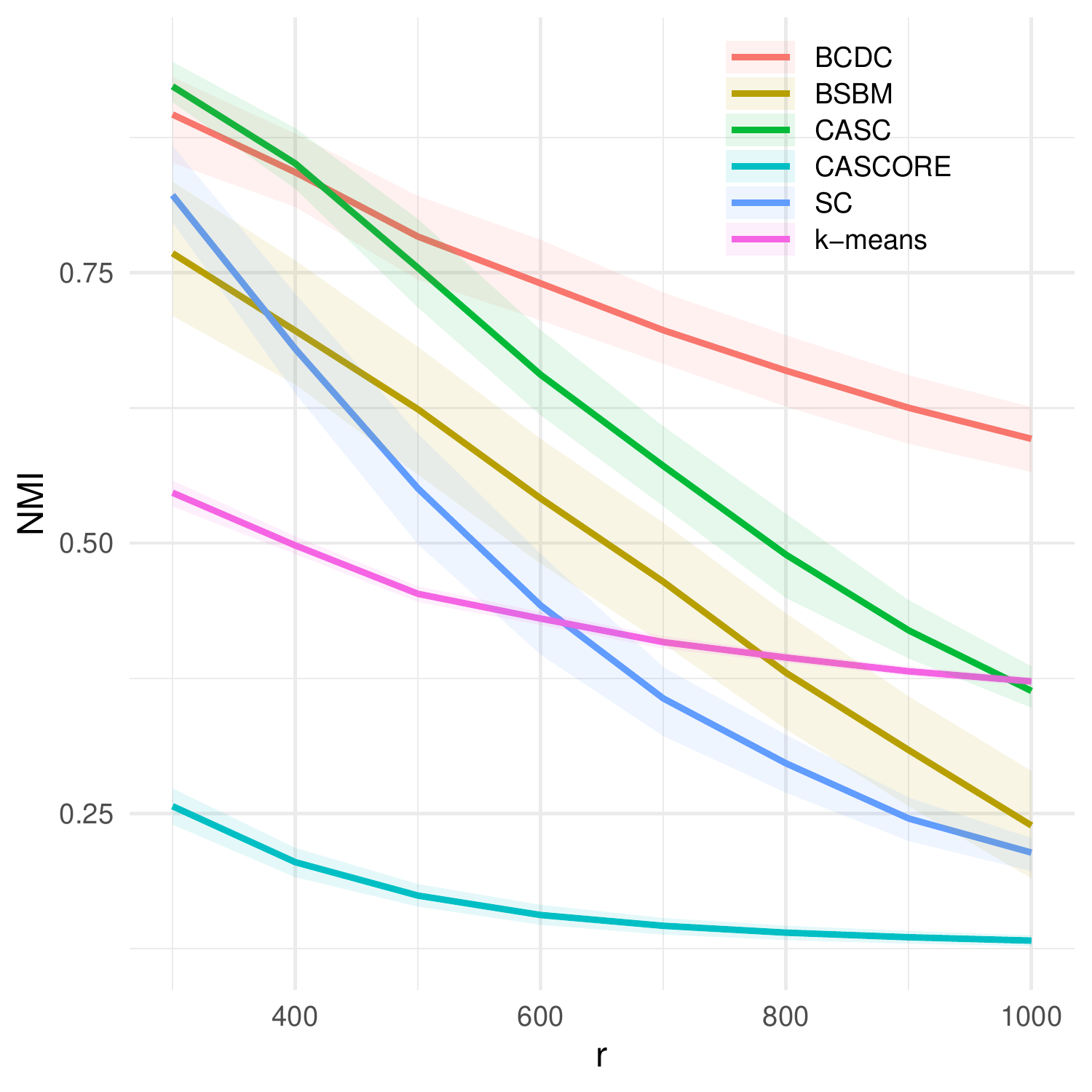}
	\includegraphics[width=.49\textwidth]{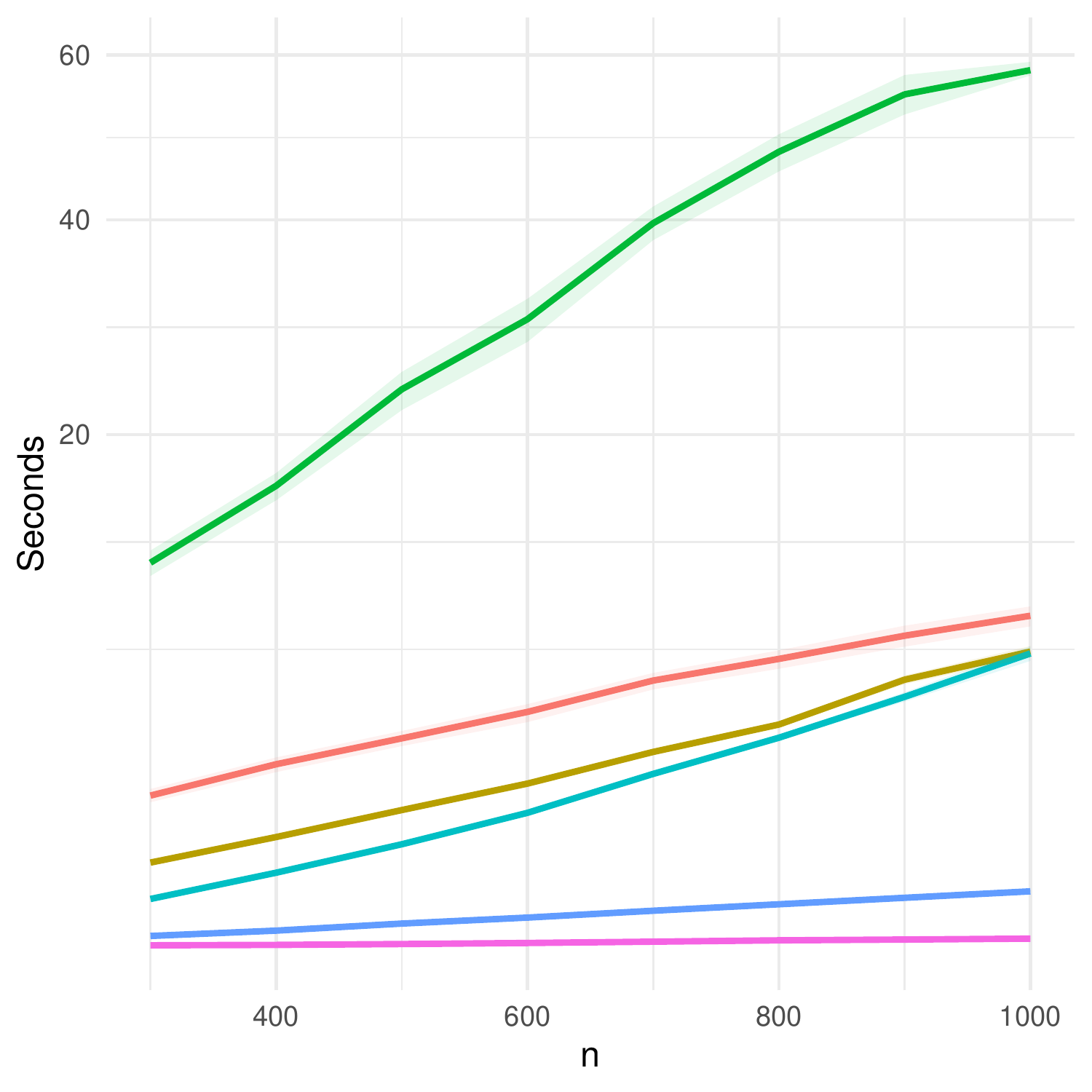}
    \caption{NMI (left) and run time (right) for five different methods when the network has a mix of continuous and discrete covariates.}
    \label{fig:boxplot_mix}
\end{figure}

The results are shown in Figure~\ref{fig:boxplot_mix}.
We find that BCDC is competitive with the other methods when $n \leq 600$, but is superior for larger networks with $n > 600$.
We also show in Figure~\ref{fig:boxplot_mix} the run time for each method.
We see that BCDC scales linearly with $n$ and is faster than CASC for all of the networks.

\subsection{Sparse networks and high-dimensional features}

We next consider a setting from~\cite{sarkar}, who proposed a covariate-regularized procedure for community detection in sparse graphs.
This allows us to explore whether our model works for sparse networks, as well as networks with high-dimensional features.
We consider the exact simulation setting as in~\cite{sarkar}, in which the networks are generated from a 3-block SBM on 800 nodes with block-size ratios $3:4:5$.
The true connectivity matrix is
\begin{equation*}
    B = 0.01
    \begin{bmatrix}
	1.6 & 1.2 & 0.16\\
	1.2 & 1.6 & 0.02\\
	0.16 & 0.02 & 1.2\\
	\end{bmatrix} \enskip ,
\end{equation*}
leading to a very sparse network, with expected average degree $\approx 5.8$.
The covariates are generated from 100-dimensional Gaussian distributions $N(\mu_{z_i}, I_{100})$, with centers that are only non-zero on the first two dimensions: 
\begin{align*}
    \mu_1=(0,2,\bm 0_{98}),\quad \mu_2=(-1,-0.8,\bm 0_{98}),\quad \mu_3=(1,-0.8,\bm 0_{98})\enskip .
\end{align*}
In this setting, it is difficult to distinguish clusters 1 and 2 using the network information alone, and clusters 2 and 3 based on nodal covariates alone.

\begin{figure}
	\centering
	\includegraphics[width=.49\textwidth]{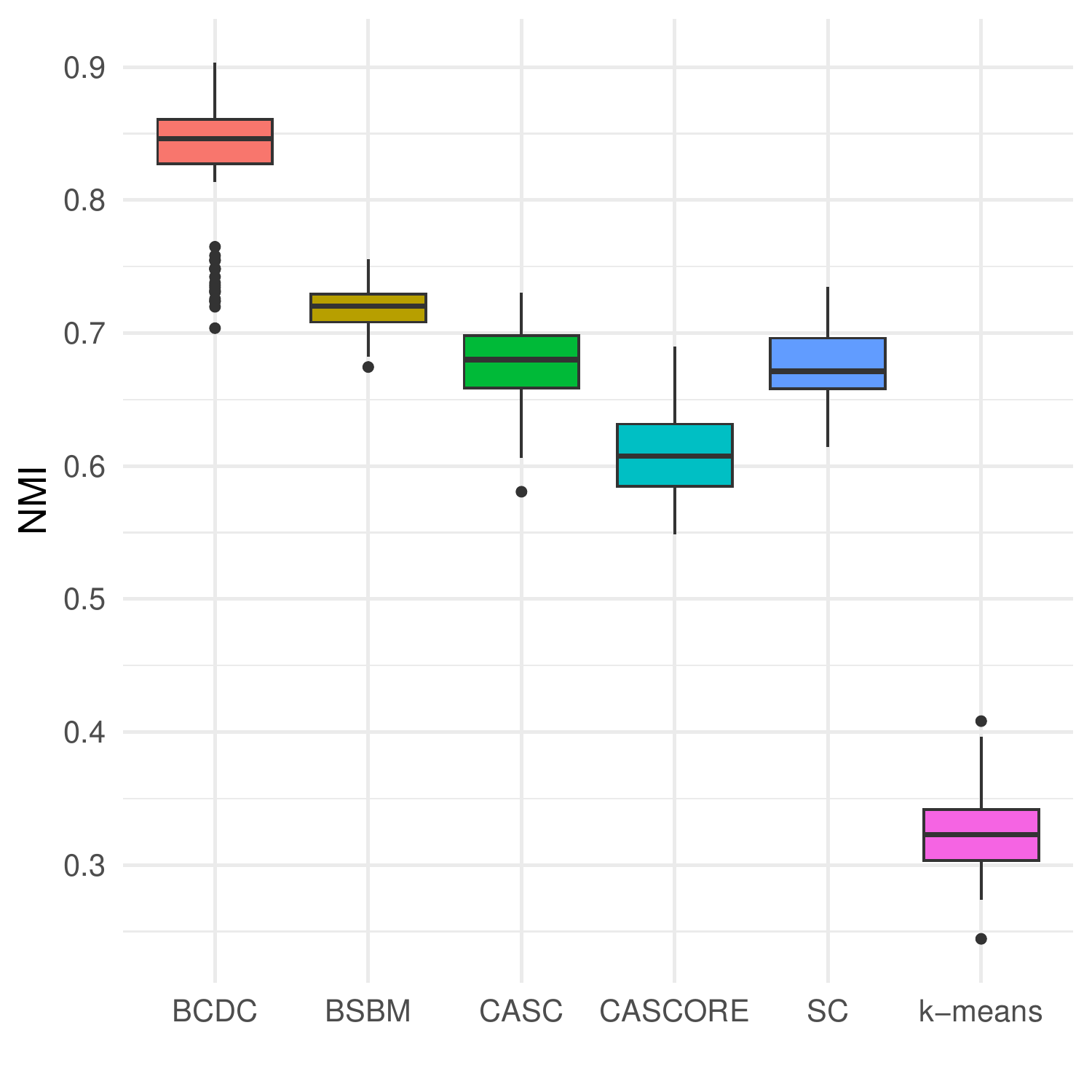}
	\includegraphics[width=.49\textwidth]{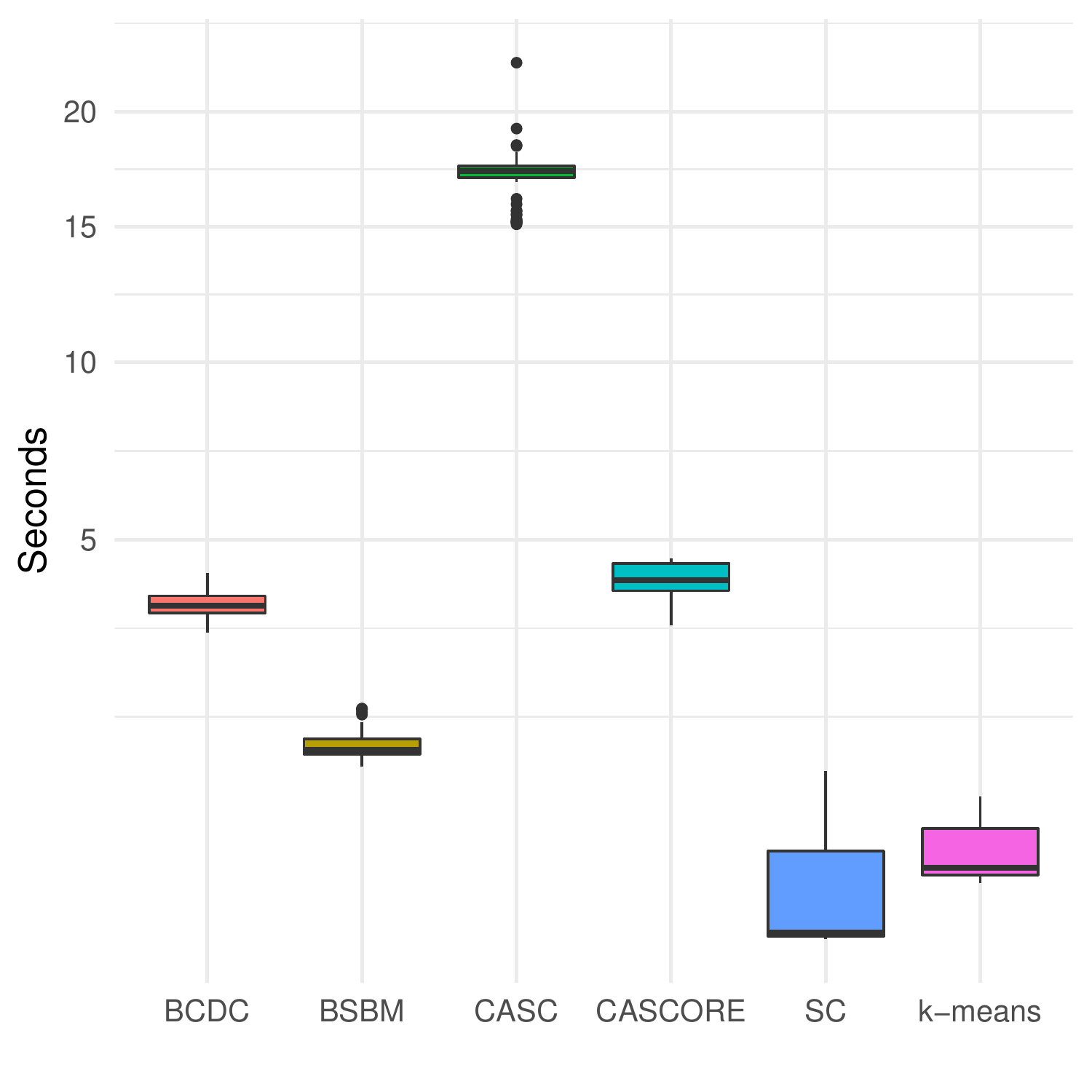}
	\caption{Boxplots of NMI (left) and run time (right) for five different methods when the network is sparse and the covariates are high-dimensional.}
	\label{fig:boxplot_sparse}
\end{figure}

This experiment was repeated 100 times for independently generated samples.
In each replicate, we ran the chain for 1,000 iterations.
The results are shown in 
Figure~\ref{fig:boxplot_sparse}.
Note that our method consistently achieves a better NMI than all of the other methods.
Although we did not carry out a direct comparison with the method from \cite{sarkar} since their work focuses on regularizing high-dimensional features, our NMI results are stable and seem to be higher than those reported in~\cite{sarkar}.
Importantly, we also see in Figure~\ref{fig:boxplot_sparse} that BCDC is only slightly slower than all of the methods that use only the network or only the covariates, but is considerably faster than CASC and CASCORE, which also use both the network and the nodal information.
This illustrates the disadvantage of CASC for larger sparse networks and highlights the efficiency of our MCMC algorithm. That CASC slows down for larger networks can be attributed to the addition of the dense $\tau XX^T$ to the sparse Laplacian $L$, resulting in an overall dense similarity matrix $L_x$.

\subsection{Homophily}

Finally, we consider a network model with homophily, which is the tendency for nodes to be connected when they share a nodal feature.
For this, we sample a categorical covariate $x_i$ with two levels, and let
\begin{equation*}
    \mathbb{P}(g_{ij} = 1 \given z_i, z_j, x_i, x_j) = P_{z_iz_j} + \beta\bm 1 \{x_i = x_j\} \enskip ,
\end{equation*}
where $P_{z_iz_j}$ is an SBM with connectivity \eqref{eq:pp:eta}, setting $p = 0.3$ and $r = 0.7$.
This creates $2K$ communities, and separates the effect of community structure from the effect of node-level covariates.
We take $K = 3$ and vary $\beta$ in $[-0.2,0.2]$.
Note that when $\beta > 0$, we say the nodes exhibit positive homophily.
We run this for $n = 600$ and $n = 1200$.

\begin{figure}[t]
    \includegraphics[width=.49\textwidth]{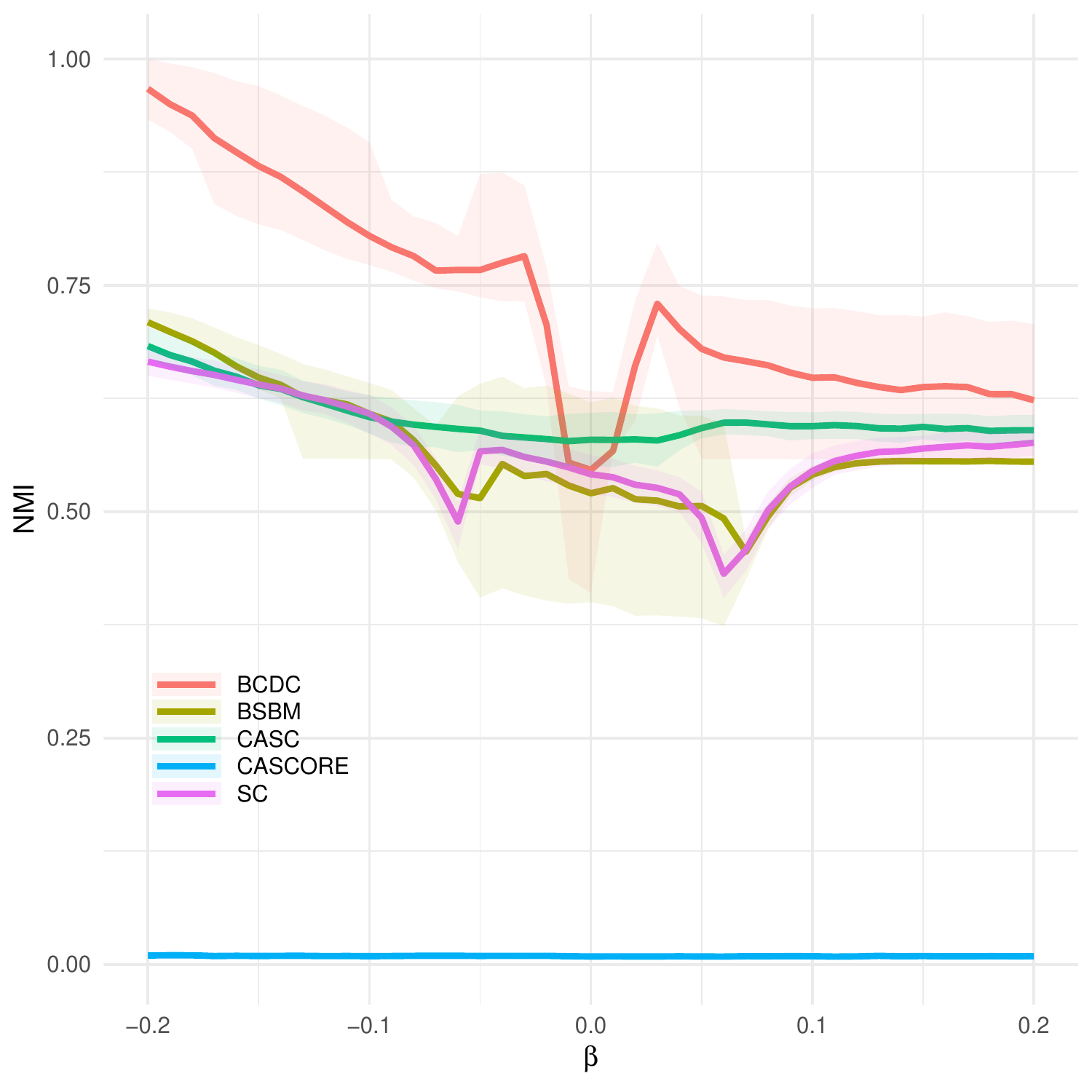}
	\includegraphics[width=.49\textwidth]{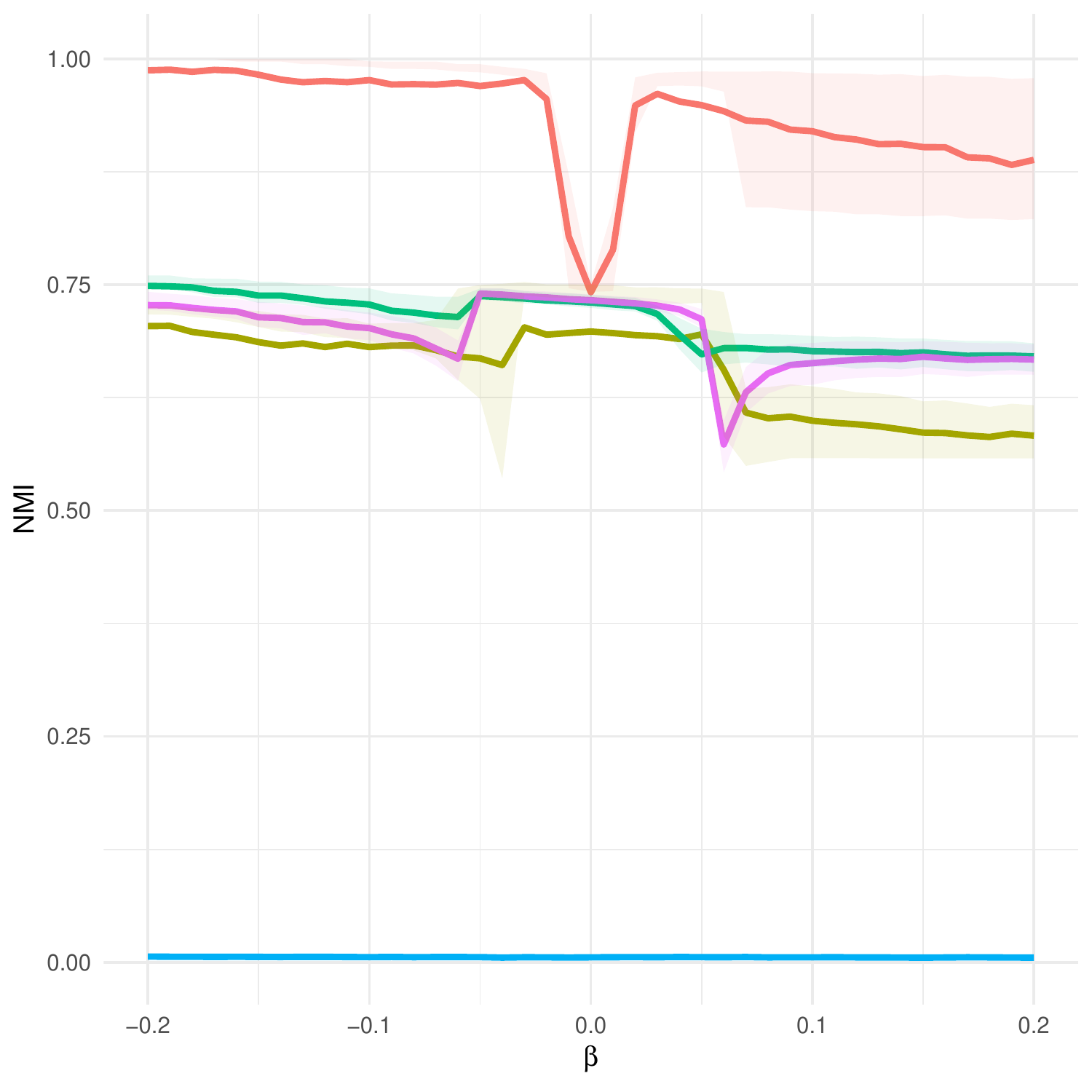}
    \caption{NMI results 
    on a 3-block SBM with a homophily effect for $n=600$ (left) and $n=1200$ (right).}
    \label{fig:boxplot_homophily}
\end{figure}

The results are show in Figure~\ref{fig:boxplot_homophily}.
We find that BCDC has superior performance to the other methods even when they are provided the true number ($2K$) of communities.
The performance increases as the homophily effect increases in magnitude, which we should expect because the homophily effect is an informative covariate that is not included with BSBM.

\section{Real data analysis} \label{sec:data}

In this section, we apply our model to the same two datasets from \cite{sarkar}, a network representing Mexican political elites and a network representing the Weddell Sea ecosystem.
We compare our the results from our models against several methods that use only the network, only the covariates, and both the network and covariates.

\subsection{Performance measures}

In addition to computing the NMI with the (alleged) ground truth labels, it is also helpful to compare the performance using the Bayesian information criterion (BIC) based on an SBM likelihood conditional on the labels.
This is especially important because, unlike in the simulations, the ``true" clusters are exogenously specified.
The (conditional) BIC is defined as the log-marginal likelihood muliplied by $-2$.
That is,
\begin{align}
    \text{BIC}(\zb) &= -2 \log \int p(A \mid \bm\eta, \pib, \zb) \,d \bm\eta\, d \pib \label{eq:bic:exact}\\
    &\approx -2  \log  p(A \mid \bm{\hat \eta}, \bm{\hat \pi}, \zb)
    + c(K) \log \binom{n}2  \enskip ,
     \label{eq:bic:approx}
\end{align}
where $K$ is the number of communities in $\bm z$, $c(K) = \frac12 K(K+1) + (K-1)$ is the degrees of freedom in the parameters $(\bm\eta, \bm\pi)$,  $\bm \pi$ is the label prior, and $(\bm{\hat \eta}, \bm{\hat \pi})$ is the maximum likelihood estimator of those parameters, i.e., the maximizer of $(\bm \eta, \bm \pi)~\mapsto~p(A~\mid~\bm\eta, \pib, \zb)$.
We assume a uniform prior over $\bm \eta$ and $\bm \pi$.
Note that~\eqref{eq:bic:approx} is the well-known approximation to the BIC \citep{schwarz1978estimating, konishi2008information} and it shows the usefulness of $\text{BIC}(\bm z)$ as a measure of performance for real networks: Due to the presence of the complexity term $\approx c(K) \log (n^2)$, we get a good balance of the model fit and the number of communities.
Label vectors $\bm z$ with smaller $\text{BIC}(\bm z)$ are thus more desirable from a block modeling standpoint, regardless of their relation to the ground truth. 

We have
\begin{align*}
    p(A \mid \bm\eta, \pib, \zb) = \prod_{k \le \ell} \eta_{k\ell}^{M_{k\ell}} (1-\eta_{k\ell})^{N_{k\ell} - M_{k\ell} } \prod_{k} \pi_{k}^{n_k(\zb)} \enskip ,
\end{align*}
where $n_k(\zb) = \sum_i 1\{z_i = k\}$, and $M_{k\ell}$ and $N_{k\ell}$ are as in~\eqref{eq:M:N:counts}. Hence, the exact BIC in our setting is
\begin{align*}
    \text{BIC}(\bm z) = -2 \Bigl[ \sum_{k \le \ell} \log B(M_{k\ell}+1,N_{k\ell} - M_{k\ell}+1) + \log \bm B(\bm n(z)+\bm 1_K) \Bigr] \enskip ,
\end{align*}
where $\bm n (\bm z) = (n_k(\bm z))$ and $\bm B(\cdot)$ is the multivariate Beta function.


\subsection{Mexican Political Elites}
	
\begin{figure}[t]
	\centering
	\includegraphics[width=0.65\textwidth]{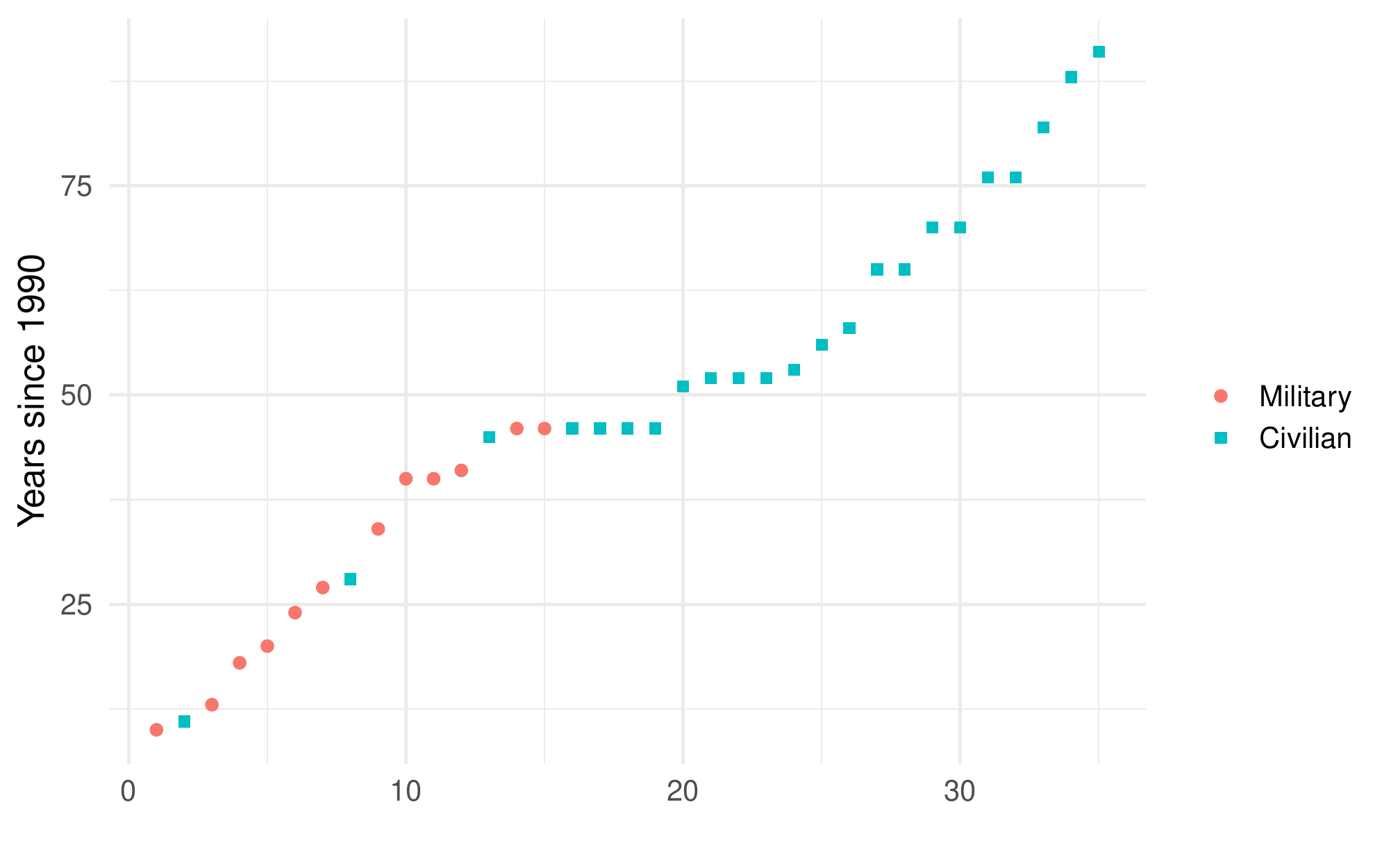}
	\caption{Node feature for the Mexican political network, which is the number of years since 1990 that the actor first got a significant governmental position.}
	\label{fig:mex_node_feature}
\end{figure}

The first dataset we consider involves Mexican political elites \citep{gil1996political}.
In this network, the $n = 35$ vertices represent Mexican presidents and their close collaborators, and the 117 edges represent significant political, kinship, friendship, or business ties among them.
The ground truth is a classification of the politicians according to their professional background: military and civilians.
The covariate we include is the number of years since 1990 that the actor first got a significant governmental position.
Figure \ref{fig:mex_node_feature} reveals that this covariate has some discriminatory power in the cluster labels.
This is due to the fact that after the Mexican revolution at the beginning of the twentieth century, the political elite was dominated by the military, and later the civilians gradually succeeded the power.

\begin{figure}[t]
    \centering
    \includegraphics[width=.42\textwidth]{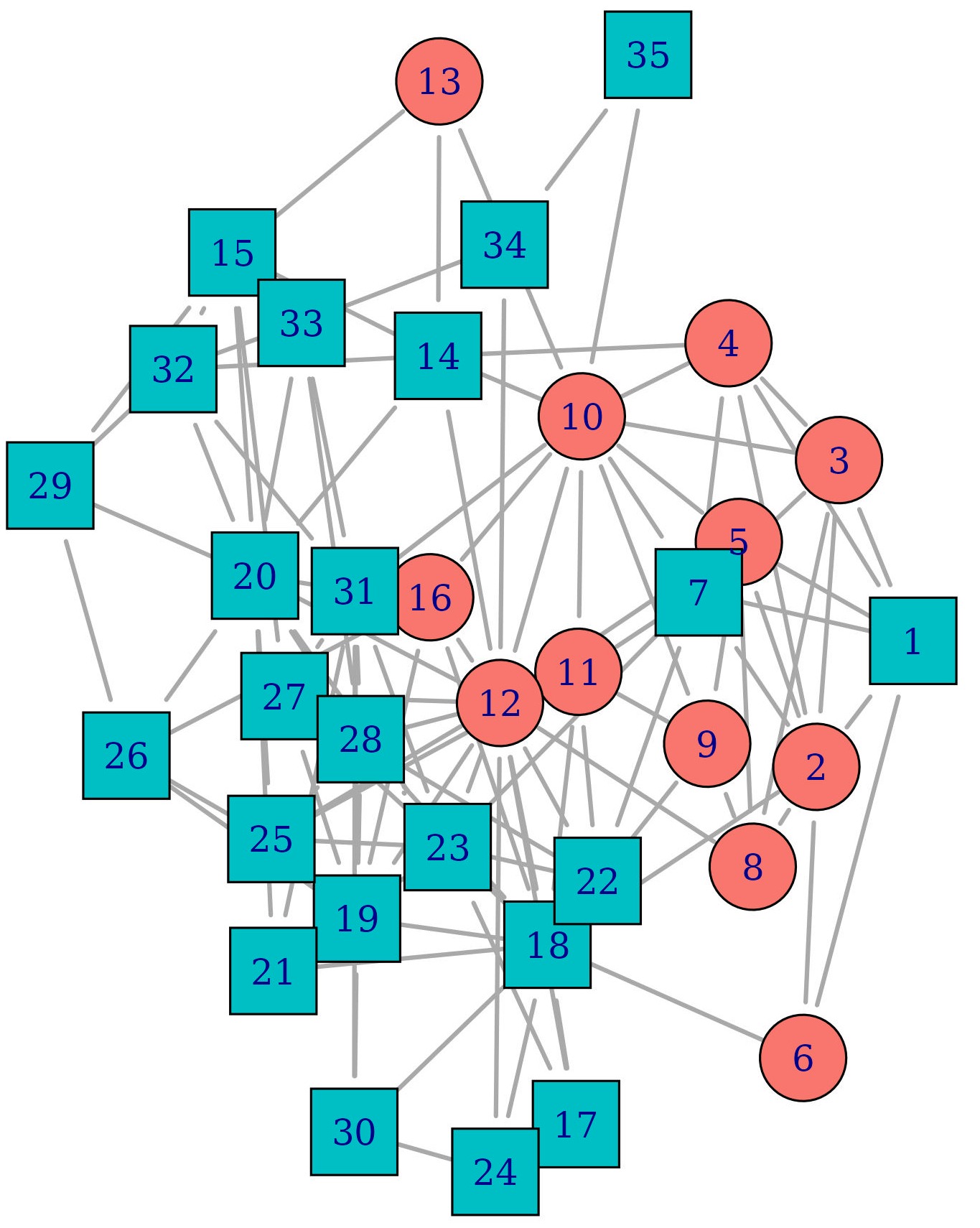}\quad
    \includegraphics[width=.42\textwidth]{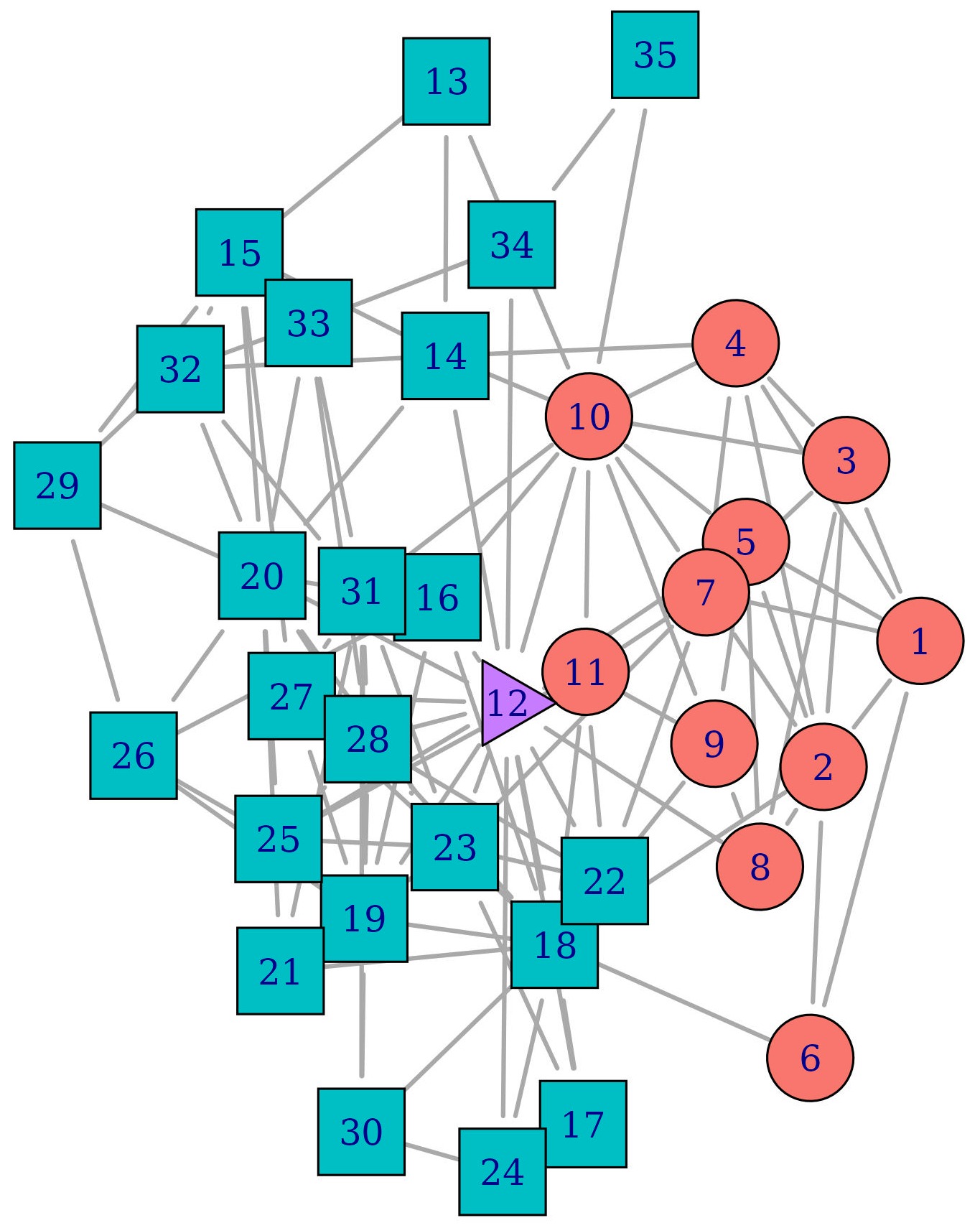}
    \caption{Mexican political network, 
    colored by true (left) and estimated (right) clusters.} 
    \label{fig:mexican_political_network}
\end{figure}

Table \ref{tab:nmi} contains the NMI results of our method compared with the same comparison methods in the simulation section.
We see that our method achieves the best results, and we visualize our estimated clusters in the network compared with the true labels in Figure \ref{fig:mexican_political_network}.
Again, for the other methods, we assume the knowledge of the true number of clusters while ours learns the number of the clusters via posterior inference for NMI comparisons.

As already pointed out in \cite{sarkar}, node 35 has exactly one connection to each of the military and civilian groups, but obtained a governmental position in the 90s, which greatly hinted at a civilian background.
By using the covariate, our method accurately captures this label.
On the other hand, node 9 seized power in 1940 when the government was almost equally represented by civilian and military politicians, which makes detecting his group difficult, but has more edges to the military group than the civilian group.
In this case, our method correctly assigns the military label to it by considering the graph structure.

\begin{table}[t]
\centering
\begin{tabular}{lcccccc}
\toprule
\textbf{Dataset} & 
\bcdc & \casc & \cascore & \kmeans & \spec & \bsbm 
\\ \midrule
Mexican politicians & \textbf{0.43} & 0.37 & 0.28 & 0.26 & 0.37 & 0.10 \\ \midrule
Weddell Sea & \textbf{0.44} & 0.25 & 0.15 & 0.35 & 0.33 & 0.23 \\ \bottomrule
\end{tabular}

\medskip
\caption{NMI results on the two real datasets.}
\label{tab:nmi}
\end{table}
	
We also notice that node 1 has five connections to the military and only one connection to the civilian, and node 1 seized power in 1911.
Similarly for node 7, which has five connections to the military and three connections to the civilian and seized power in 1928.
For these nodes, both the network and the covariates strongly indicate a closer relationship to the military, which is what our method assigns despite the true label showing civilian.
Finally, our method assigns node 12 to its own cluster.
This is likely because this is the highest-degree node with 5 military connections and 12 civilian connections.
However, in researching node 12, we discovered that Miguel Alem\'an Vald\'es was the first civilian president after several military presidents, which suggests that there may have been a labeling error in the original publication of this dataset from \cite{gil1996political}.
This could explain why our method has the best BIC, even better than the ``true" labels, in Table~\ref{tab:bic}.

\begin{table}[t]
\centering
\begin{tabular}{p{1.8cm}ccccccc}
\toprule
\textbf{Dataset} & ``\true" & \bcdc & \casc & \cascore & \kmeans & \spec & \bsbm \\ \midrule
Mexican politicians & 636 & \textbf{586} & 587 & 607 & 626 & 587 & 598 \\ \midrule
Weddell Sea & 138k & \textbf{33k} & 124k & 119k & 144k & 100k & 71k \\ \bottomrule
\end{tabular}

\medskip
\caption{BIC results on the two real datasets.}
\label{tab:bic}
\end{table}

\subsection{Weddell Sea Ecosystem}

\begin{figure}
    \centering
    \includegraphics[width=.49\textwidth, frame]{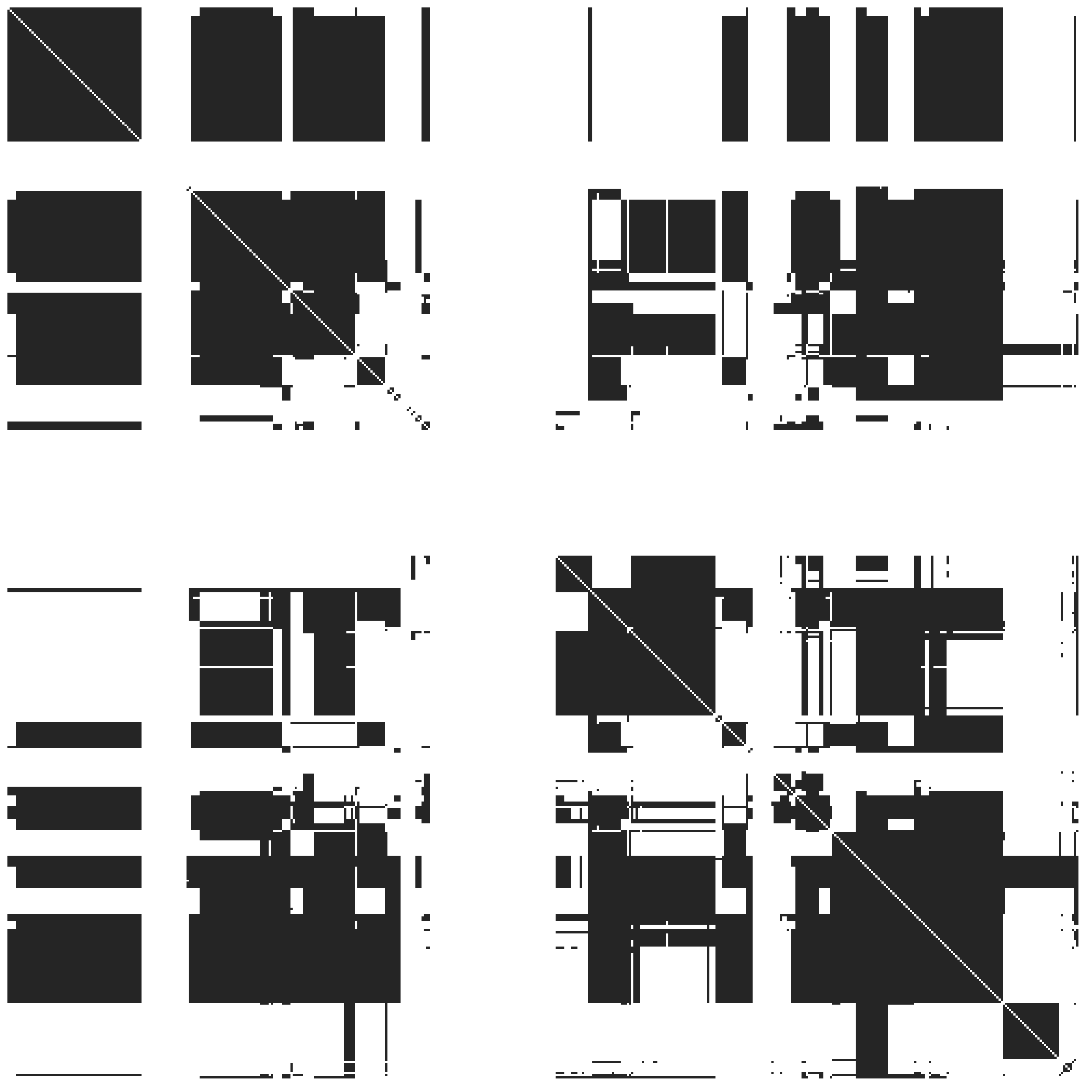}
    \includegraphics[width=.49\textwidth, frame]{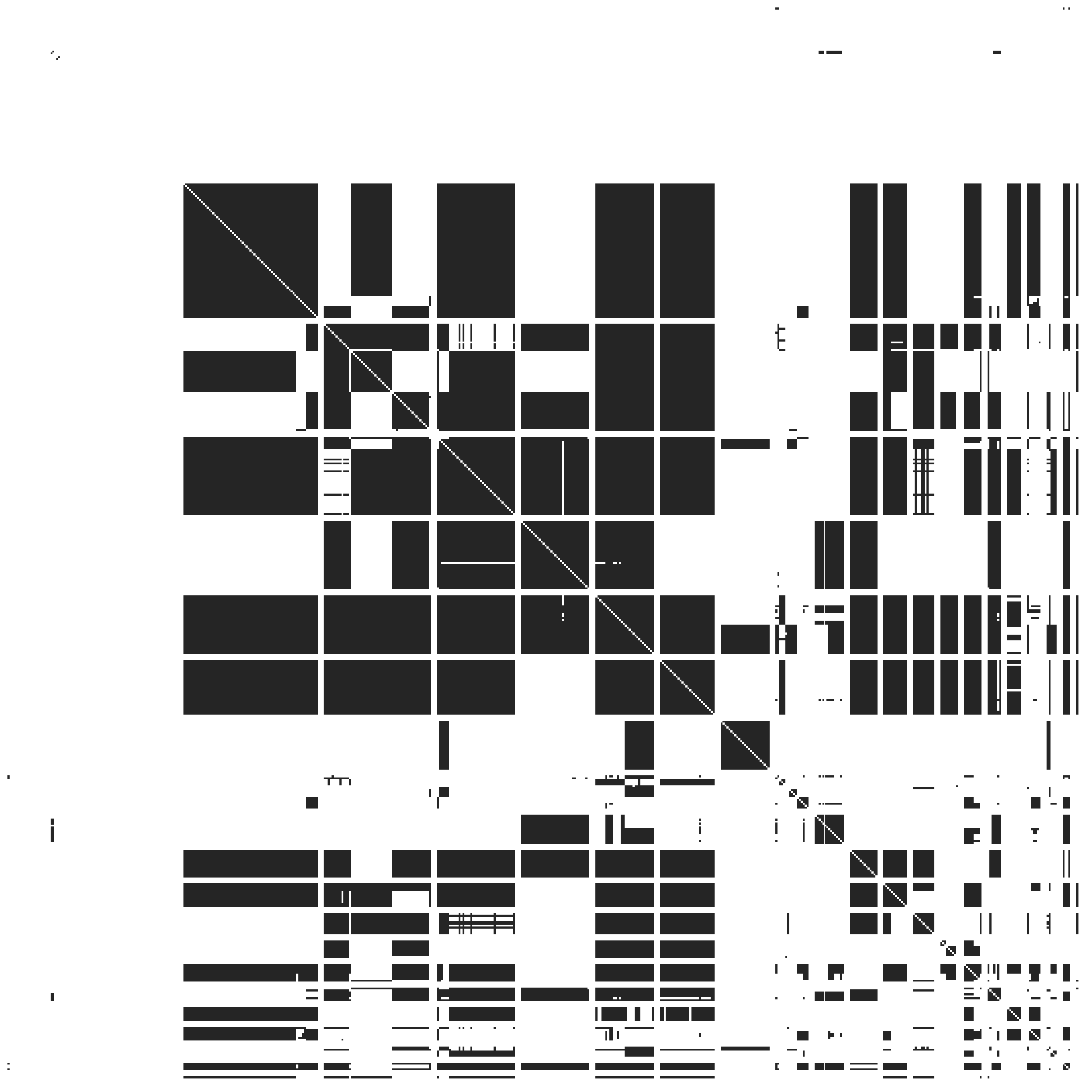}
    \caption{Visualization of the adjacency matrix for the Weddell Sea network clustered by the true feeding type (left) and estimated clusters (right). In each case, the rows and columns of the matrix are permutated so that nodes in the same cluster form contiguous blocks. Clusters on the right are ordered according to their size.}
    \label{fig:weddell_network}
\end{figure}

The second dataset we consider is a predator-prey, directed network representing the marine food web of the Weddell Sea off of the Antarctic Peninsula, which was collected by \cite{jacob2011role}.
Since ecosystems are complex, interconnected environments, network analyses have emerged as a popular technique for untangling these connections.
The Weddell Sea network has 487 nodes that signify different marine species, and there is a link between nodes $i$ and $j$ if species $i$ (predator) feeds on species $j$ (prey).
Following \cite{sarkar}, we construct a binary, undirected network from this directed network in which $A_{ij} = 1$ if there are at least 5 common prey between species $i$ and $j$, and $A_{ij} = 0$ otherwise.
The network is shown in Figure~\ref{fig:weddell_network}.

In \cite{jacob2011role}, the authors analyze the relationship between the body size of each species and its feeding type: primary producer, herbivorous/detrivorous, detrivorous, carnivorous, carnivorous/necrovorous, and omnivorous.
Figure \ref{fig:bodymass} shows these body sizes grouped by feeding type, where, again following \cite{sarkar}, we group detrivorous, carnivorous, carnivorous/necrovorous as ``Carnivore" to obtain four groups.
The adjacency matrix in Figure~\ref{fig:weddell_network} (left) is also sorted by these groups.
The authors of \cite{jacob2011role} found body size to be positively correlated with trophic level, but noted that ``predators on intermediate trophic levels do not necessarily feed on smaller or prey similar in size but depending on their foraging strategy have a wider prey size range available."
Therefore, body size is insufficient on its own to distinguish the groups, and it would be preferable to consider the interconnectedness of the food web when tasked with clustering the species.

Table \ref{tab:nmi} shows that BCDC provides the best clustering results compared to the other methods.
Therefore, using all of the available information provides an improvement in clustering accuracy over the use of just the network structure or the nodal information.
As before, BCDC is the only method that did not know that there are four ``true" groups.
Interestingly, BCDC estimates many more clusters -- 21 in total -- which may explain its higher NMI, since we see qualitatively in Figure~\ref{fig:weddell_network} (right) a more refined block structure.
This is quantified and corroborated through BIC in Table~\ref{tab:bic}, which again shows our method outperforms even the ``true" clusters.
All of this suggests there may be distinct sub-blocks within the Herbivore, Carnivore, and Omnivore classes.

\section{Discussion}\label{sec:discussion}

In this work, we proposed a Bayesian model for community detection in networks with covariates in which both the network and node features of the network are jointly utilized for estimating community structure.
In particular, the contribution of nodal information is explicitly modeled in the prior distribution for the community labels via a covariate-dependent random partition prior.
We proposed efficient MCMC algorithms for sampling the posterior distributions of all the parameters including the community labels and the number of the communities.
Numerical studies demonstrated the overall superior performance of our model over many of the existing methods.

Compared to an almost exclusive literature of frequentist methods, our work is among the first in proposing a Bayesian approach for tackling the problem, which confers some notable advantages in terms of uncertainty quantification, as well as estimating all the model parameters.
Notably, unlike the other methods in the literature, our model estimates the number of communities via posterior inference without any knowledge or prior information on the true number.
Future work will be devoted to developing Bayesian models for community detection in degree-corrected SBMs and dynamic network models.

We can also easily extend our model to a partially-observed SBM in the spirit of \cite{zhou2015infinite}.
Specifically, we can modify~\eqref{eq:SBM} to 
\begin{align*}
    P(A \given \etab, \zb)=\prod_{1\leq i<j\leq n} \Bigl[ \eta_{z_i,z_j}^{A_{ij}}(1-\eta_{z_i,z_j})^{1-A_{ij}}\Bigr]^{\mask_{ij}} \enskip ,
\end{align*}
where $\maskb = (m_{ij}) \in \{0,1\}^{n \times n}$ is a (symmetric) observation mask, with $m_{ij} = 1$ for the observed edges.
This only effects sampling $\zb$ and $\beta$ through the modified counts
\begin{align*} 
	O_{i\ell} = \sum_{j: j \neq i} \mask_{ij} A_{ij} 1\{z_j = \ell\}, \quad
	n_{i\ell} = \sum_{j: j\neq i} \mask_{ij} 1\{z_j = \ell\} \enskip ,
\end{align*}
replacing~\eqref{eq:O:n:def}, and
\begin{align*}
	M_{k\ell} = \sum_{(i,j) \in \Gamma_{k\ell}} 
	\mask_{ij}A_{ij} 1\{z_i = k, z_j = \ell\},
	\quad
	N_{k \ell} =  \sum_{(i,j) \in \Gamma_{k\ell}} \mask_{ij}1\{z_i = k, z_j = \ell\} \enskip ,
\end{align*}
replacing~\eqref{eq:M:N:counts}. This allows 
our model to also predict missing edges.

Finally, it may be of interest to test whether there is an association between the node covariates and inferred community structure.
One approach to this is with Bayes factors comparing models with and without covariates, using, for example, the approach in \cite{legramanti2020bayesian} for testing partition structures in SBMs.
While this is a principled Bayesian approach, it only tests whether the set of covariates provides a more parsimonious clustering than without the covariates rather than identifying which covariates are significant.
One idea for testing individual covariate significance is to test for a difference between the posterior distributions of the cluster centers $\xib$ implied by $\zb$.
Note that this corresponds to testing whether the priors $\nu$ on auxiliary probability distributions $q$ have overlapping variances, but we leave a rigorous treatment of this idea to future work.


\bibliographystyle{ba}
\bibliography{refs}

\begin{acks}[Acknowledgments]
We are very grateful to the Editor, the Associate Editor and two reviewers for their valuable comments.  LS and LL were supported by NSF grants DMS 2113642 and DMS 1654579. AA would like to acknowledge the support of NSF grant DMS-1945667. NJ was partially supported by NIH/NICHD grant 1DP2HD091799-01.
\end{acks}

%
%
%
%
%
\end{document}